\documentclass[preprint]{aastex}

\slugcomment{Accepted for publication in ApJ}

\shorttitle{Outflows in Extreme NLS1s}
\shortauthors{Leighly \& Moore}

\begin{document}


\title{HST STIS Ultraviolet Spectral Evidence of Outflow in Extreme
Narrow-line Seyfert 1 Galaxies: I. Data and Analysis\footnote{Based on observations made
with the NASA/ESA Hubble Space Telescope, obtained at the Space
Telescope Science Institute, which is operated by the Association of
Universities for Research in Astronomy, Inc., under NASA contract NAS
5-26555. These observations are associated with proposal
\#7360.}\hphantom{l}{$^,$}\footnote{Based on observations obtained at Cerro Tololo
Inter-American Observatory, a division of the National Optical
Astronomy Observatories, which is operated by the Association of
Universities for Research in Astronomy, Inc. under cooperative
agreement with the National Science Foundation}}


\author{Karen M. Leighly and John R. Moore}
\affil{Department of Physics and Astronomy, The University of
Oklahoma, 440 W. Brooks St., Norman, OK 73019}
\email{leighly@nhn.ou.edu, jmoore@nhn.ou.edu}



\begin{abstract}
We present {\it HST} STIS observations of two extreme NLS1s,
IRAS~13224$-$3809 and 1H~0707$-$495.  The spectra are characterized by
very blue continua, broad, strongly blueshifted high-ionization lines
(including \ion{C}{4} and \ion{N}{5}), and narrow, symmetric
intermediate- (including \ion{C}{3}], \ion{Si}{3}], \ion{Al}{3}) and
low-ionization (e.g., \ion{Mg}{2}) lines centered at their rest wavelengths.
The emission-line profiles suggest that the high-ionization lines are
produced in a wind, and the intermediate- and low-ionization lines are
produced in low-velocity gas associated with the accretion disk or
base of the wind.  In this paper, we present the analysis of the
spectra from these two objects; in a companion paper we present
photoionization analysis and a toy dynamical model for the wind.  The
highly asymmetric profile of \ion{C}{4} suggests that it is dominated
by emission from the wind, so we develop a template for the wind from
the \ion{C}{4} line. We model the bright emission lines in the spectra
using a combination of this template, and a narrow, symmetric line
centered at the rest wavelength.  We also analyzed a comparison sample
of {\it HST} spectra from 14 additional NLS1s, and construct a
correlation matrix of emission line and continuum properties.  A
number of strong correlations were observed, including several 
involving the asymmetry of the \ion{C}{4} line.  
\end{abstract}




\section{Introduction}

In 1992, it was demonstrated by Boroson \& Green that the optical
emission line properties in the region of the spectrum around H$\beta$
are strongly correlated with one another.  A principal components
analysis allowed the largest differences among optical emission line
properties to be gathered together in a construct commonly known 
as ``Eigenvector 1''.  The strongest differences hinged on the
strength of the \ion{Fe}{2} and [\ion{O}{3}] emission, and the
width and asymmetry of H$\beta$.  This set of strong correlations is 
remarkable, as it involves correlations among the dynamics and gas
properties between emission regions separated by vast distances.
Furthermore, the same set of correlations are observed in samples
selected in many different ways.  This pervasiveness leads us to
believe that we are observing the manifestation of a primary physical
parameter. A favored explanation is that it is the accretion
rate relative to the black hole mass onto the active nucleus (e.g.,
Boller, Brandt \& Fink 1996; Pounds, Done \& Osborne 1995; 
Comastri et al.\ 1998; Leighly 1999ab;  Mathur 2000;  Mineshige et al.\ 2000; 
Kuraszkiewicz et al. 2000; Pounds et al.\ 2001; Puchnarewicz et al.\
2001; Ballantyne, Iwasawa, \& Fabian 2001; Boroson 2002; Crenshaw,
Kraemer \& Gable 2003). 

Narrow-line Seyfert 1 galaxies (NLS1s\footnote{In this paper, the term
NLS1s is used to refer to all objects fulfilling the optical criteria
regardless of their luminosity.  Likewise, the term broad-line Seyfert
1 is used to refer to objects of both Seyfert and quasar
luminosities.}) are identified by their optical emission line
properties.  They typically have narrow permitted optical lines (FWHM
of H$\beta<2000\rm\, km\,s^{-1}$), weak forbidden lines
([\ion{O}{3}]/H$\beta<3$; this distinguishes them from Seyfert 2
galaxies), and frequently they show strong \ion{Fe}{2} emission
(Osterbrock \& Pogge 1985; Goodrich 1989).  These are the same
properties that define the Boroson \& Green (1992) Eigenvector 1, and
indeed, NLS1s fall at one end of Eigenvector 1.  Thus, the study of
NLS1s offers an attractive research opportunity: if we can understand
the origin of the emission-line properties of NLS1s, then we may be in
a position to understand AGN emission in general.  Furthermore,
because as least some of the lines in NLS1s are narrow, identification
of the frequently strongly-blended lines is less ambiguous than it is
in broad-line objects.

NLS1 research dramatically increased after the publication of a
seminal paper in 1996 by Boller, Brandt \& Fink describing their
discovery that NLS1s have systematically steeper {\it ROSAT} spectra
than AGN with broad optical lines.  Subsequently it was
shown from {\it ASCA} observations that the hard X-ray spectrum is
also steeper (Brandt, Mathur \& Elvis 1997; Leighly 1999b), and that the
X-ray emission shows higher fractional amplitude of variability
(Leighly 1999a).  These results were generally interpreted as evidence
that the black hole is systematically smaller in NLS1s, implying a
higher accretion rate relative to the Eddington value compared with
Seyfert 1 galaxies with broad optical lines.

Once the high accretion rate paradigm was established, it could be
used to understand why NLS1s have narrow optical lines.
Boller, Brandt, \& Fink (1996) mention that if the black hole
mass in NLS1s is smaller, and the emission lines are produced at the same
radius as in broad-line objects (due to the same radiating flux)  the
dynamical velocities will be smaller, and narrower lines will be
produced.  This theme was expanded upon by Wandel \& Boller (1998),
who consider the fact that the steeper soft X-ray spectrum of NLS1s, when
extrapolated into the UV, implies a larger ionizing flux relative to
their luminosity.   This causes the emission region to be larger where the
velocities are  smaller. 

However, the optical emission lines are only one side of the coin; UV
emission lines, generally characterized by a higher degree of
ionization, may also provide valuable information.  This is because
the optical and UV emission lines are generally not produced in the
same gas.  The best evidence for this comes from so-called
reverberation mapping experiments (e.g., Peterson 1993).  Also, the
profiles of the emission lines suggest this as well, as the higher
ionization lines are observed to be generally broader than the lower
ionization lines (e.g., Baldwin 1997).  Therefore, we may learn more
about NLS1s by studying their UV emission lines.

Several groups have investigated the UV emission line properties of
NLS1s.  The first paper investigating a sample of IUE spectra from
NLS1s was published by Rodr\'iguez-Pascual, Mas-Hesse \&
Santos-Lle\'o (1997).  They reported that the strongest UV emission
lines have broad wings, unlike the optical emission lines, and they
speculate that the UV line-emitting gas may be optically thin in the
continuum.

Kuraszkiewicz et al.\ (2000) present a detailed study of {\it IUE} and
{\it HST} spectra from NLS1s.  They found that, compared with
broad-line quasars, the equivalent widths of \ion{C}{4} and
\ion{Mg}{2} are low.  They found that, in the cases where the
lines could be deblended, the \ion{Si}{3}]/\ion{C}{3}] ratio was high.  In
addition, the \ion{Si}{4}+\ion{O}{4}] feature near 1400 \AA\ was
found to be strong compared with \ion{C}{4}, and the \ion{Al}{3}
equivalent width was high.  They also found that although the UV
emission lines in NLS1s are broader than the Balmer lines, they still
tend to be narrower than those lines in broad-line quasars.  They then
model many of the emission line ratios using a 1-zone {\it Cloudy}
model (Ferland 2001).  They infer that the line ratios imply that the
ionization parameter is lower and the density is higher in NLS1s than
in broad-line Seyfert 1 galaxies.  We comment on their inferences in
the companion paper to this one (Leighly 2004a; hereafter referred to
as Paper II).  

Wills et al.\ (1999), studying the {\it HST} spectra from a small
sample of PG quasars, found that Eigenvector 1 is also associated with
a high \ion{Si}{3}] to \ion{C}{3}] ratio, stronger UV low-ionization
lines, weaker \ion{C}{4}, but stronger \ion{N}{5}.  They infer from
these spectra that NLS1s are typified by high densities, low
ionization parameters, and nitrogen enhancement, perhaps from nuclear
starbursts. 

All the work done so far on the UV spectra of NLS1s has been generally
confined to studies of the properties of the integrated lines, to a
greater or lesser extent, depending on the analysis.  Specifically,
the conditions of a single-zone BLR are modelled or discussed more or
less quantitatively, and evidence for a radially-extended BLR is
inferred when particular lines are found to not be consistent with the
one-zone model (e.g., Kuraszkiewicz et al.\ (2000) infer that, since
the observed \ion{Mg}{2} is underpredicted by the one-zone model, it
must be produced in a different region than the other UV emission
lines).  However, we know that generally the emission line profiles
are not all the same, and higher ionization lines tend to have broader
profiles.  This implies that the gas properties are different among
regions with larger or smaller line-of-sight velocities, and therefore
a study of the average properties could be misleading.

In this paper, we present the observations and analysis of {\it HST}
observations of two NLS1s that have truly remarkable UV spectra. In
Section 2, we introduce the targets, describe our observations and
discuss the analysis of the emission lines.  In Section 3, we
compare the results from these two NLS1s with spectra from 14 NLS1s
drawn from the {\it HST} archive, and perform a correlation analysis
and a principal components analysis on the results.  In a companion
paper II, photoionization analysis, and a toy dynamical model are
presented, and the results are discussed.  Partial results from this
study have been published previously in Leighly (2000) and Leighly
(2001).  We use $H_0=50\rm\, km\,s^{-1}\,Mpc^{-1}$ and $q_0=0.5$
unless otherwise noted.

\section{Observations and Analysis}

\subsection{The Targets}

We observed two NLS1s, IRAS~13224$-$3809 (z=0.066, V=15.9) and
1H~0707$-$495 (z=0.040, V=15.7), using {\it HST} STIS during 1999.
Groundbased optical spectra were also obtained.  
The observing log is given in Table 1.

These two NLS1s are quite remarkable objects and can be considered to
be extreme for their class in terms of their X-ray properties. They
showed among the largest fractional amplitudes of variability in a
sample of NLS1s observed by {\it ASCA} (Leighly 1999a).  {\it ROSAT}
monitoring observations of IRAS~13224--3809 found strong soft X-ray
flaring by up to a factor of 60 in as short time interval as two
days (Boller et al.\ 1997).  Large amplitude X-ray variability was
also reported during a long {\it ASCA} observation (Dewangen et al.\
2002).  Their X-ray spectra are also unusual.  They show extremely
strong X-ray soft excess components and lie on the extreme of the
variability--soft excess correlation (Leighly 1999b; Leighly 2004b).
{\it XMM--Newton} observations of 1H~0707$-$495 and IRAS~13224$-$3809
reveal a peculiar deep absorption feature near 7~keV that is quite
difficult to explain (Boller et al.\ 2002; Boller et al.\ 2003).  The
{\it XMM-Newton} observation of 1H~0707$-$495 showed primarily
color-independent X-ray variability (Boller et al.\ 2002).  In
contrast, a {\it Chandra} observation of 1H~0707$-$495 found strong
spectral variability in that the soft X-rays varied much less than the
hard X-rays (Leighly et al.\ 2002; Leighly et al.\ in prep.).

The motivation for the {\it HST} observations was the discovery of a
peculiar feature near 1 keV observed in the {\it ASCA} spectra
(Leighly et al.\ 1997).  This feature appeared to be an absorption
edge, but the energy was clearly too high to be ascribed to
highly-ionized oxygen, which is generally the most important source of
X-ray opacity. Leighly et al.\ (1997) speculated that oxygen was
indeed responsible for the feature, and, if so, the absorbing gas was
outflowing at rather large velocities.  This interpretation was
strengthened by the similarity between luminous NLS1s and
low-ionization BALQSOs (e.g., Boroson \& Meyers 1992 for I~Zw~1).
Reasoning that X-ray absorption may be accompanied by UV absorption,
we proposed to make use of the much superior spectral resolution
(relative to {\it ASCA} CCDs) provided by {\it HST} STIS, and look for
evidence for outflowing gas in the UV.  We note here that no evidence
for absorption intrinsic to the quasars was found in the spectra.
This result leaves the issue of the X-ray features unresolved, since
it is certainly plausible that the X-ray absorption lines could be
produced in gas too highly ionized to present opacity in the UV.  We
also note that subsequent high resolution X-ray observations using
{\it Chandra} reveal absorption lines (Leighly et al.\ 2002; Leighly
et al.\ in prep.)

\subsection{The {\it HST} observations and Continua}

Our observations were made during the first cycle that the STIS
spectrometer was available.  We made the {\it HST} observations using
a ``pattern'' in order to decrease possible systematic error due to
fixed pattern noise.  Three or four steps along the slit were used for
each spectroscopic element.  The relative wavelength calibration of
each spectrum was checked prior to summing the spectra by cross
correlating with a preliminary average spectrum.  Many Galactic
absorption lines were observed in the spectra. These were used to
identify and correct zero-point wavelength offsets in the spectra.
Then, the redshift was obtained from the optical and UV spectra.  In
the optical spectrum of IRAS~13224$-$3809, H$\beta$ is extended in the
spatial direction.  The extended emission probably originates in a
starburst in the host galaxy, and therefore provides a good estimate
of the systemic redshift.  The extended emission contributes to the
narrowest part of the H$\beta$ profile; therefore, we estimated the
redshift from centroid of the top of the H$\beta$ line.
Interestingly, \ion{O}{3} has a blue offset of $370 \rm \,km\,s^{-1}$
(Grupe \& Leighly 2002), similar to I~Zw~1 (Phillips et al.\ 1976).
The redshift of 1H~0707$-$495 was estimated also using the centroid of
the top of the H$\beta$ line.  These redshifts were consistent with
that inferred from \ion{Mg}{2}, if that line is centered at the rest
wavelength, and the centroid of the narrow part of Ly$\alpha$.  The
measured redshifts were 0.066 and 0.040 for IRAS~13224$-$3809 and
1H~0707$-$495, respectively.

The UV spectra are displayed in Fig.\ 1 along with nonsimultaneous
optical spectra obtained at CTIO (details in Table 1) and a composite
quasar spectrum from the LBQS (Francis et al.\ 1991) for
comparison. This figure shows that IRAS~13224$-$3809 and 1H~0707$-$495
have continua as blue as the average quasar.  This result is
in agreement with the discovery by Grupe et al.\ (1998) that soft
X-ray selected AGN from the {\it ROSAT} All Sky Survey, a sample which
is comprised of 1/2 NLS1s, generally have quite blue optical--UV
continuum spectra; some are as blue as predicted by the standard
(thin) accretion disk model ($F_{\nu} \propto \nu^{1/3}$; $F_{\lambda}
\propto \lambda^{-7/3}$). For comparison, a line with a slope of
$-7/3$ is included on Fig.\ 1 (which is plotted in terms of
$F_\lambda$), as well as one also with slope $-0.32$, which is the
median slope in the LBQS (Francis et al.\ 1991).  Very blue optical
spectra seem to be common among luminous NLS1s (see also
RGB~J0044+193: Siebert et al.\  1999; RX~J0134.2$-$4258: Grupe et al.\
2000; RX~J2217.9$-$5941: Grupe,  Thomas \& Leighly 2001; PHL 1811:
Leighly et al.\ 2001). 

\begin{figure}
\plotone{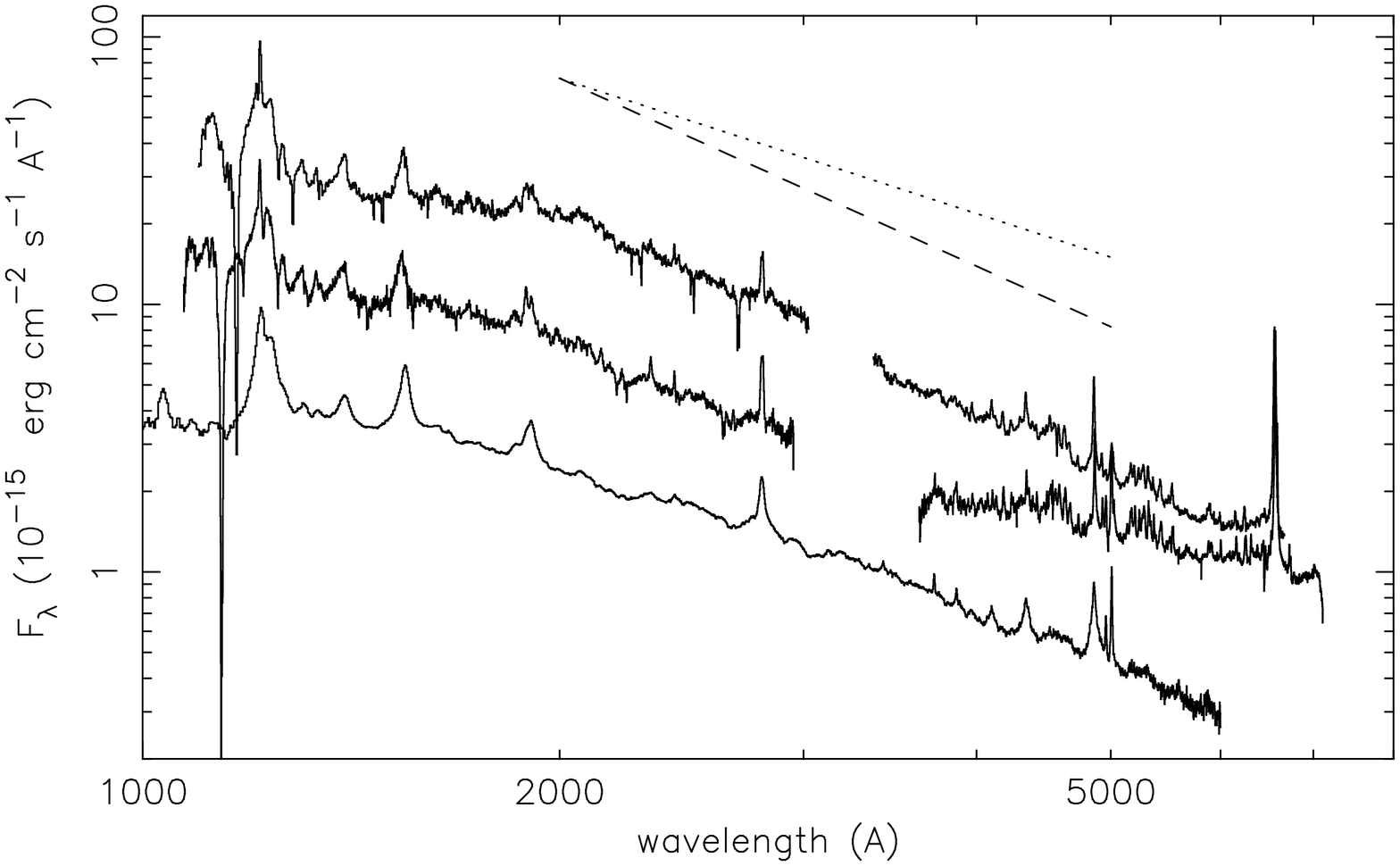}
\caption{The broadband continua of 1H~0707$-$495 (top) and
  IRAS~13224$-$3809 (middle) compared with that of a composite quasar
  spectrum  (bottom; Francis et al.\ 1991). The dashed line is a power
  law with slope of $-7/3$, which is the slope predicted by the
  standard thin accretion disk model.  The dotted line is a power law with
  slope $-0.32$, which is the median in the LBQS (Francis et al.\
  1991). \label{fig1}}  
\end{figure}

\placefigure{fig1} 

The optical spectrum from 1H~0707$-$495 appears to be slightly offset
in normalization from the UV spectra.  This offset could be due to 
aperture effects, or from variability, as there was a one-year
interval between the optical and UV spectra.  However, the continuum
slopes in the UV and optical spectra appear to be consistent.  In
contrast, the optical spectrum from IRAS~13224$-$3809 appears to be
much flatter than the UV continuum spectrum.  This object is an
extremely luminous far-infrared source (Norris et al.\ 1990), and we
suspect that the flattening is due to contamination of the central
engine emission by the starburst.  We suspect that similar
contamination was present in other spectra from this object, leading
to reports that the optical--UV continuum spectrum is flat (e.g.,
Boller et al.\ 1993).  The UV spectra presented here show that the
central engine continuum spectrum is not flat, but very clearly blue.

\subsection{The Emission Lines and Modeling}

The spectra are shown in more detail in Fig.\ 2.  In this plot, we
follow Laor et al.\ (1997) and mark the rest wavelengths of many of the
strong lines frequently observed in AGN spectra.  Our identification
of the emission features are discussed below.

\begin{figure}
\epsscale{1.0}
\plotone{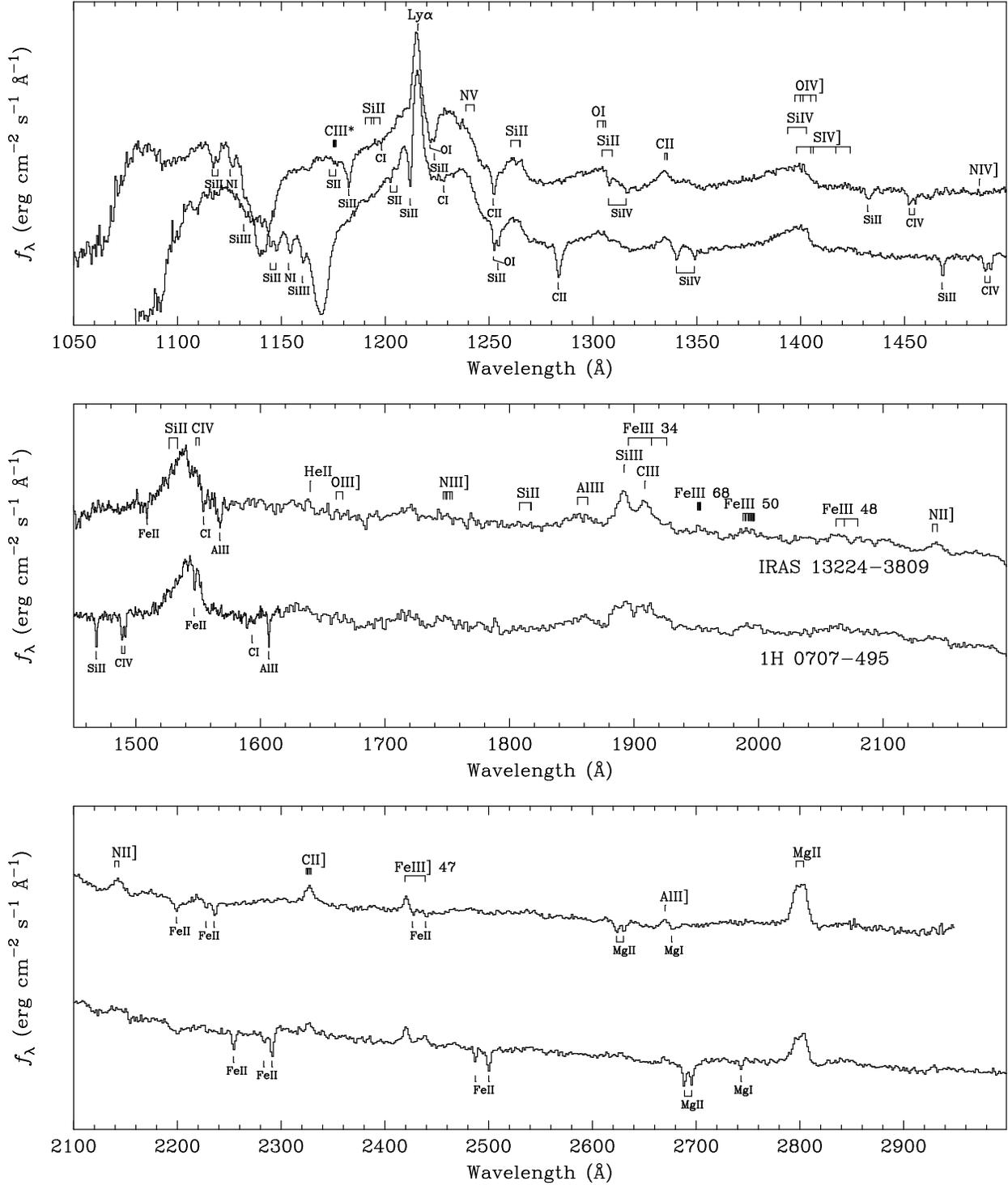}
\caption{The {\it HST} STIS spectra from IRAS 13224$-$3809 and
  1H~0707$-$495, shifted to the rest frame and dereddened as described
  in the text.  For clarity, spectra have been renormalized to have
  the same flux, and offset from one another.  The rest frame
  wavelengths of strong emission lines frequently seen in AGNs are
  labeled, excluding the many \ion{Fe}{2} multiplets.  All identified
  absorption lines are labeled; they all originate in our
  Galaxy.\label{fig2}}\end{figure}

\placefigure{fig2}

These spectra are characterized by the following properties:
\begin{itemize}
\item Strong \ion{Fe}{2} emission.  Emission lines from \ion{Fe}{3}
are also identified.  The \ion{Fe}{2} is strong enough that it forms a
pseudocontinuum makes identification of the real continuum difficult,
but it can also be identified as a forest of relatively
narrow lines, especially near 2000 \AA\/.
\item Broad and blueshifted high-ionization lines. The high-ionization
lines, including Ly$\alpha$, \ion{N}{5}, the \ion{Si}{4}--\ion{O}{4}]
complex, and \ion{C}{4} are all relatively broad (FWHM are 
approximately 5000$\,\rm km\,s^{-1}$) and blueshifted, with little
emission redward of the rest wavelength, at least in \ion{C}{4} and
\ion{N}{5}.
\item Relatively narrow intermediate- and low-ionization lines
  centered at the rest wavelength. \ion{C}{3}], \ion{Si}{3}] and
  \ion{Mg}{2} are narrow  (900--1900$\rm \,km\,s^{-1}$), and their
  profiles are reflection-symmetric about about the rest wavelength.
\item Strong lower ionization lines, including \ion{Si}{2} and
\ion{C}{2}.  
\end{itemize}

The spectra most strongly resemble that of the NLS1 prototype I~Zw~1,
which also shows blueshifted high-ionization lines, narrow
intermediate- and low-ionization lines and strong \ion{Fe}{2} and
\ion{Fe}{3} (Laor et  al.\ 1997).  However, as discussed in
Section~3, they are somewhat 
different than the other narrow-line Seyfert 1 galaxies and quasars.
The presence of strong low-ionization lines in our spectra is similar
to that seen in several PG quasars (Wills et al.\ 1999; see their
Figure 1). 

The profound differences in the profiles of the high- and
low-ionization lines are highlighted in Fig.\ 3, which compares the
profile of the prototypical high-ionization line \ion{C}{4} with the
profile of a low-ionization line \ion{Mg}{2}, for our NLS1s and for a
composite quasar spectrum\footnote{Note that composite spectra
constructed on the basis of redshifts measured from a single line may
have broader-than-average and more-symmetric-than-average lines.}
(Francis et al.\ 1991).  It has been previously noted that
high-ionization lines are frequently broader and strongly blueshifted
compared with the low-ionization lines (e.g., Tytler \& Fan 1992;
Marziani et al.\ 1996).  The velocity offsets in \ion{C}{4} described
by Marziani et al.\ (1996) range from zero to $\sim 3000$; the
centroids of our profiles are offset by $\sim 2500 \rm \, km\,
s^{-1}$, as large as the largest values values found by Marziani et
al.  These profiles are most simply explained if the high-ionization
line-emitting gas is accelerated toward us in a wind, while the
low-ionization line-emitting gas is located on the surface of the
accretion disk, or in the low-velocity base of the wind.  The wind may
be accelerated by radiative-line driving from the strong UV continuum.
This process is commonly inferred to be occurring in hot stars and
cataclysmic variables, but it has also been applied to AGN (e.g.,
Murray et al.\ 1995; Murray \& Chiang 1998; Proga, Stone, \& Kallman
2000), and we discuss this further in Paper II.  The accretion disk is
assumed optically thick, so that we do not see the receding wind on
the other side.  Hence, the emission lines are predominantly
blueshifted.  Collin-Souffrin et al.\ (1988) were the first to suggest
a two-component broad-line region in which the high-ionization lines
are produced in a wind, and the low-ionization lines are associated
with the outer parts of the accretion disk.  The spectra presented
here arguably provide the strongest support for this scenario to date.

\begin{figure}
\epsscale{1.0}
\plotone{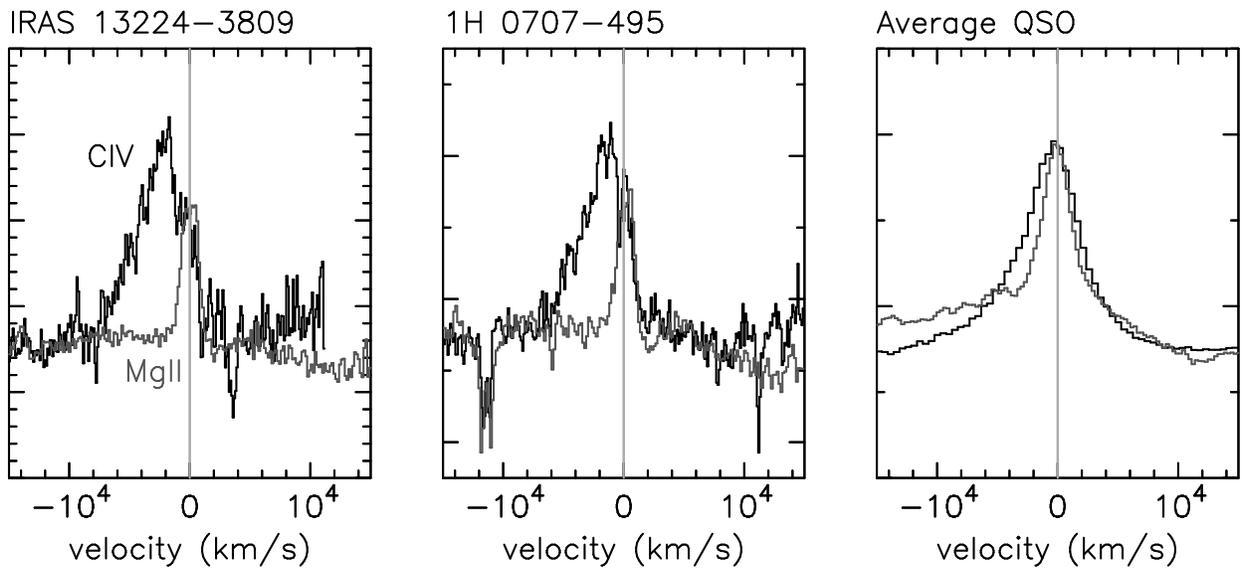}
\caption{The rescaled representative high-ionization line
\ion{C}{4} superimposed on the representative low-ionization line
\ion{Mg}{2} as a function of velocity, for our two NLS1s and the
average quasar (Francis et al.\ 1991).  The average quasar
high-ionization line is slightly broader and slightly blueshifted
compared with the low-ionization line.  In contrast, the
low-ionization lines in NLS1s are much narrower and the
high-ionization lines are strongly blueshifted.\label{fig3}}
\end{figure}

\placefigure{fig3}

Because the high-ionization lines are nearly disjoint in velocity
space compared with the low- and intermediate-ionization lines, we
model them separately. We then use the results to constrain
photoionization models for the conditions of the disk and wind
separately in Paper II\footnote{We do not know with certainty the
geometrical and physical origin of the emission lines in the objects
we are discussing here.  However, for simplicity, we refer to the
highly blueshifted high-ionization lines as originating in the
``wind'', and the narrow, symmetric low-ionization lines as
originating in the ``disk''. These distinctions are somewhat similar
to the HIL and LIL regions previously proposed (Collin-Souffrin et
al.\ 1988), except that \ion{C}{3}] and other similar
intermediate-ionization lines in our spectra appear to be produced in
the disk.}.  Such an undertaking is fraught with peril.  In the
emission-line profile, we observe only the velocity component parallel
to our line of sight.  Especially problematic is the gas at zero
velocity, as the emitting matter may have huge transverse velocity
that we cannot detect.  However, such an approach is not without
precedent: Baldwin et al.\ (1996) analyzed the spectra from a luminous
quasar Q0207$-$398 in this way.

To measure the line properties, we adopted a modified version of the
procedure used by Baldwin et al.\ (1996).  We outline the procedure
here, and describe the application to each region of the spectra
below; the results are given in Fig.\ 4 and  Table 2. First, we make
several simplifying assumptions.  Fig.\ 3 suggests that, to zeroth
order, the wind and disk lines are kinematically disjoint.  We further
assume, to begin with, that all high-ionization lines have the same
velocity profile; we comment on the validity of that assumption below.
Then, since \ion{C}{4} is isolated compared with the other
high-ionization lines, we use it to create a wind-profile template
that is used to model the other high-ionization lines.  The procedure
used to develop the template is described in the next section.

The intermediate- and low-ionization lines appear to be predominately
symmetric about their rest wavelengths; therefore, we model them using
symmetric Lorenzian or Gaussian profiles.  Preliminary examination of
spectra showed that we could not assume the same velocity
width for all the intermediate- and low-ionization lines, as
\ion{Mg}{2} could be seen to be clearly significantly narrower than
\ion{C}{3}].  Therefore, in fitting the intermediate- and
low-ionization lines, we allowed the width to remain an adjustable
parameter so long as it could be constrained usefully in the fitting.
Generally, the widths could be usefully constrained unless the lines
are very weak (e.g., Section 2.3.7), or unless the feature
is comprised of several heavily-blended components, in which case we
fixed the velocity width for particular individual lines to the values
measured from other lines.

The next sections explain the process that we used to model the
spectra.  Throughout, we use a locally-defined continuum.  This is
necessary because the strong \ion{Fe}{2} and \ion{Fe}{3} multiplets
that may form a pseudo-continuum that masks the real continuum.  We note
that placement of the continuum is a potentially significant source of
uncertainty in the measurement of the emission-line flux.

\subsubsection{The \ion{C}{4} region}

First, the continuum was locally identified around the \ion{C}{4}
emission line, then fit with a linear model and subtracted.  Next, we
removed the portions of the profile contaminated by absorption lines
originating in our Galaxy (Blades et al.\ 1988).  The Galactic column
is significant in the direction of our quasars, and there are a number
of fairly strong lines apparent in the spectrum.  Careful examination
of the \ion{C}{4} profile reveals faint absorption lines that were
detected by Blades et al.\ but were not identified.  The regions of
the profile contaminated by absorption lines were removed and the
profile was fitted with a high-order spline model.

The \ion{C}{4} line is a doublet, but the final template should
consist of the contribution of only one component.  To derive the
profile of one component, we used a procedure similar to that used by
Baldwin et al.\ 1996.  We assumed that each of the two components
contributed equally to the line (that is, the line is optically
thick).  The approximate contribution of one component of the doublet
was then obtained from the spline model by a sort of bootstrap method.
The doublet components are separated by 4--5 wavelength bins.
Therefore, the 4 or 5 points at longer wavelength side of the spline
model consists solely of the lower energy component, so these can be
used to determine the amplitude of the 4 or 5 longest wavelength
points of the template.  Points shortward of this consist of the sum
of both lower-energy and the higher-energy components.  At the next
point, where $i=6$, for example, the amplitude of the higher-energy
component at $i$ is already known, because it must have the amplitude
of the lower energy component at $i=$1--2.  Then, since the sum of
the lower and higher-energy components at any point $i$ must equal the
spline model at that point, the amplitude of lower-energy profile at
point $i$ can be solved for.  Thus, the amplitudes of the remaining
points could be determined iteratively.  The \ion{C}{4} emission
lines, the contribution of each multiplet and the resulting profiles
are shown in Fig.\ 4.  Interestingly, the shape of the template is
slightly but distinguishably different between the two objects.

\begin{figure}
\epsscale{1.0}
\plotone{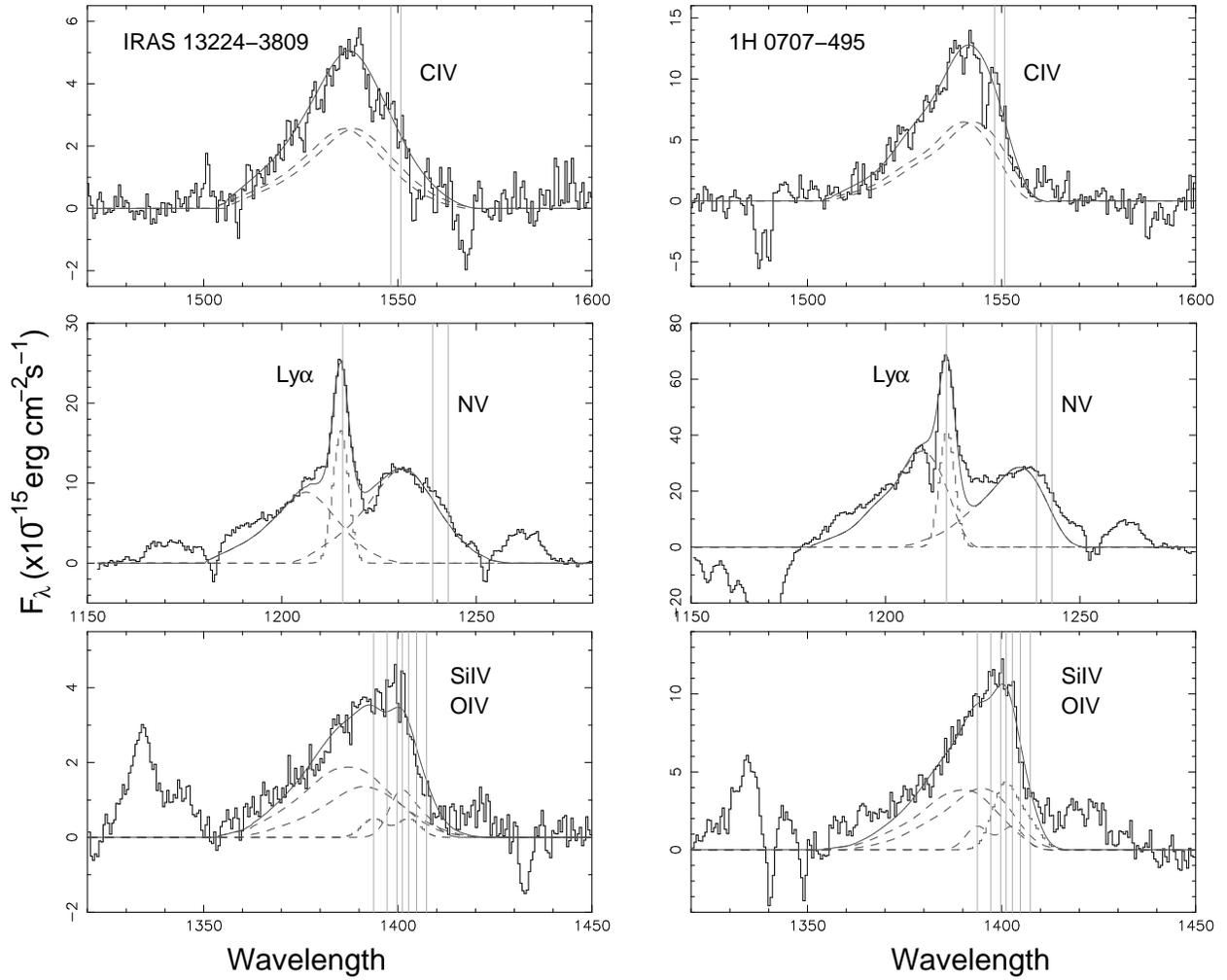}
\caption{Line models for the \ion{C}{4} region, the
Ly$\alpha$/\ion{N}{5} region, and the \ion{Si}{4}/\ion{O}{4}] region.
The dark grey dashed lines show the components and the light grey
vertical lines show the positions of the lines in the rest frame.
Resonance lines were assumed to be optically thick and thus their
ratios were assumed to be 1:1.  \ion{O}{4}] was assumed to be
optically thin (see text).\label{fig4}}
\end{figure}

\begin{figure}
\epsscale{1.0}
\plotone{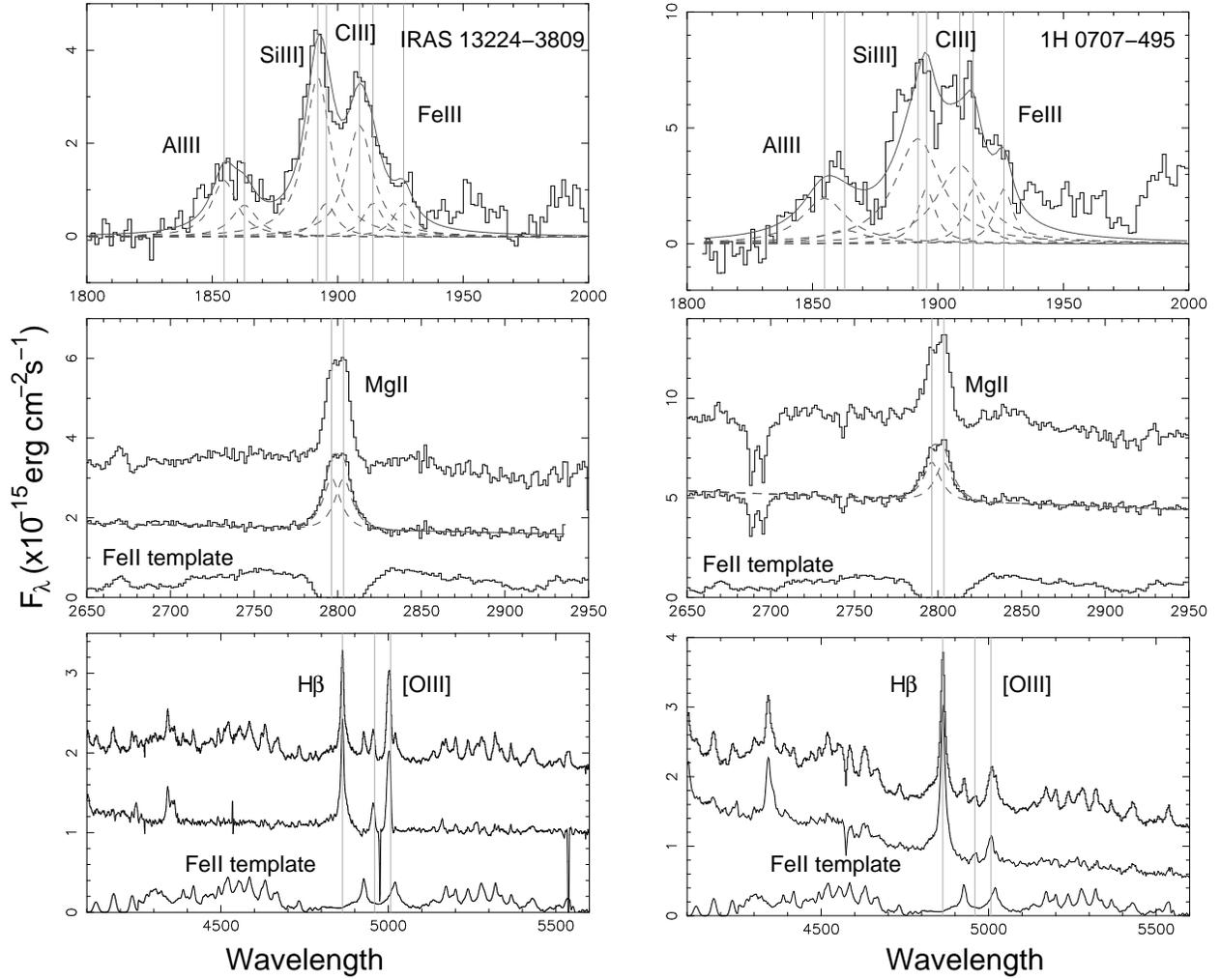}
\caption{Line models for the
\ion{Al}{3}/\ion{Si}{3}]/\ion{C}{3}] region, the \ion{Mg}{2} region,
and the H$\beta$/[\ion{O}{3}] region.  The dark grey dashed lines show
the components and the light grey vertical lines show the positions of
the lines in the rest frame.\label{fig5}}
\end{figure}

\placefigure{fig4}
\placefigure{fig5}

\subsubsection{The Ly$\alpha$ and \ion{N}{5} region}

The template developed using the \ion{C}{4} profile, shown in Fig.\ 4,
was next fit to the Ly$\alpha$/\ion{N}{5} region in both spectra.  A
second-order polynomial was used to model the local continuum, which
was then subtracted.  The fitting was done by first transforming the
template spectrum to the rest energy of the emission line in question
and then adjusting the normalization until some part of the profile
did not obviously exceed the height of the spectrum; i.e., if the
model were subtracted from the data, no obvious negative residuals
would be present.  A narrow component of Ly$\alpha$ was also clearly
visible. This is inferred to originate from the disk, and it is
modelled as a Gaussian.  Both components of the \ion{N}{5} doublet
were modeled and the flux ratio between the doublet components was
again assumed to be 1 to 1.  The results are shown in Fig.\ 4.

This procedure resulted in a good fit for IRAS~13224$-$3809.  Excess
emission shortward of Ly$\alpha$ may come from \ion{C}{3}* $\lambda
1176$ line, a line that has been identified in the spectrum of the
prototype NLS1 I~Zw~1 (Laor et al.\ 1997).  \ion{Si}{2} is strong
elsewhere in the spectrum and may contribute near 1190 Angstroms as
well.  Note that the deficit near 1222 \AA\ in the rest frame is
caused by Galactic absorption (Fig.\ 2).

The fit results for 1H~0707$-$495 are less satisfactory.  Excess
emission is present both redward and blueward of the \ion{N}{5} line.
The blueward emission could also be present in IRAS~13224$-$3809 but
it may be hidden by the Galactic absorption line near 1222
\AA\/.  This excess emission could be an indication that our
assumption that all of the resonance lines have the same profile is to
some degree incorrect; however, we cannot clearly see the blue
side of that line because it is blended with the broad Ly$\alpha$
line, so we cannot clearly evaluate whether this assumption is good or
not.  We plan to address this question in a future paper involving the
\ion{O}{6} line observed in the {\it FUSE} spectrum of 1H~0707$-$495.

In both cases, since the wind profile fits \ion{N}{5} well, we
conclude there is no disk contribution to this line.  Both a disk and
a wind component are necessary, on the other hand, to model Ly$\alpha$.

\subsubsection{The \ion{Si}{4}/\ion{O}{4}] region}

The feature near 1400\AA\ is clearly broad and asymmetric and at least
part of it must be emitted in the wind.  This is a complicated feature
in AGN in general, because it may be comprised of emission from both
\ion{O}{4}] and \ion{Si}{4}.  \ion{Si}{4} is usually a
lower-ionization line than the other lines produced in the wind.
The disk contributes low-ionization lines (e.g., \ion{Mg}{2}, as in
Fig.\ 3); therefore it is possible that there is a contribution
from \ion{Si}{4} from the disk in the 1400 \AA\/ feature.  To try to
account for this possibility in a general way, our model for this
region consists of components of wind emission from each of \ion{Si}{4}
and \ion{O}{4}] that are modelled using the template developed from
\ion{C}{4} as discussed in Section 2.3.1.  We also include
components of disk emission from each of \ion{Si}{4} and \ion{O}{4}]
that are modelled as Gaussians, as discussed above.  There are
thus 4 components in the model.

We assumed that the emission ratios of the two components of the
\ion{Si}{4} doublet have equal flux (optically thick).  For
\ion{O}{4}] we assumed optically thin ratios.  These depend partially
on density and were found to be 0.011:0.225:0.458:0.082:0.225 for the
$\lambda$1397.2, $\lambda$1399.8, $\lambda$1401.2, $\lambda$1404.8 \&
$\lambda$1407.4 emission lines respectively (Flower \& Nussbaumer
(1975) and Nussbaumer \& Storey (1982) assuming a density of
$10^9\rm\,cm^{-3}$.

We first determined the upper limits of the four components listed
above by overlaying each one on the line profile separately and
adjusting the amplitude until some part of the component was as large
as the line; i.e., no large negative fit residuals were permitted.
Those upper limits are listed in Table 2.   

Next, we investigated whether or not the entire feature could have
been emitted in the wind; that is, is the disk component necessary?
We overlayed the sum of the \ion{O}{4}] and \ion{Si}{4} wind
components on the profile, adjusting the amplitudes as described
above.  While the shortest wavelength, blueshifted part of the feature
could be modeled by the sum of the wind components from \ion{O}{4}] and
\ion{Si}{4}, the longest wavelength part, near the rest energies of
\ion{O}{4}] and \ion{Si}{4}, could not.  We inferred that the excess
emission near the rest energies of these features is emission from the
disk.  To model the disk component, emission line blends for
\ion{Si}{4} and \ion{O}{4}] were constructed using Gaussians, assuming
again that \ion{Si}{4} is optically thick and \ion{O}{4}] is
optically thin, as above.  The widths of these Gaussians could not be
independently determined in the fit, so we assumed a FWHM of 7
Angstroms, corresponding to $1500\, \rm km\,s^{-1}$.   This is close
to the value found for the 1900 \AA\/ lines (Section 2.3.4).

The 1400 \AA\/ feature is very complicated.  We show a model in Fig.\
4 that appears to fit the profile, and the parameters for this model
are listed in the Table 2; however, we do not claim that this model is
unique.  All we can really conclude is that emission from both the
disk and the wind is necessary to explain the profile; the wind
portion is needed to explain the shorter wavelength, blueshifted part,
and the disk portion is needed to explain the rest-wavelength part.
But we really cannot tell how much  of each component 
originates in each of \ion{Si}{4} and \ion{O}{4}].  The
photoionization modelling presented in Paper II is helpful for
understanding this feature. 

Adjacent to the 1400\AA\/ feature, there is excess emission near 1420
\AA\ that could be another emission line.  A possible candidate could
be \ion{S}{4}] $\lambda\lambda$ 1416.93, 1423.86 (Harper et al.\
1999). 

\subsubsection{The \ion{Al}{3}/\ion{C}{3}]/\ion{Si}{3}] region}

In the wavelength range between 1800 and 2000 \AA\/, we expect to find
emission from \ion{C}{3}], \ion{Si}{3}] and \ion{Al}{3}, as well as
possibly \ion{Fe}{3} and \ion{Fe}{2}.  Identification of the continuum
in this region is very difficult because of the strong \ion{Fe}{2} and
\ion{Fe}{3} lines. In IRAS~13224$-$3809 we clearly identify fairly
strong contributions from \ion{C}{3}], \ion{Si}{3}] and \ion{Al}{3}.
These lines appear to be narrow, symmetric, strongly peaked, and
centered at the emission line wavelength.  Thus, we infer they are
from the disk and that no measurable wind component is present in the
1900\AA\/ feature.  We use the spectral fitting package {\it LINER}
(Pogge \& Owen 1993) to model the lines.  We first attempt to model
the emission lines as Gaussians; however, they are too strongly peaked
for this to provide a good fit.  Therefore, we use Lorenzian profiles
and obtain a better fit, although we note that Lorenzian profiles may
overestimate the line flux due to the large contributions in the wings
that are difficult to distinguish from the continuum.  However, there
remains excess emission around 1925 \AA\ that most probably originates
in the \ion{Fe}{3} 34 multiplet.  This multiplet consist of three
lines at 1895.46, 1914.06, and 1926.30 \AA\ (e.g., Graham, Clowes \&
Campusano 1996).  We include these three emission lines in the model,
assuming that their fluxes are equal (e.g., Hartig \& Baldwin 1986;
see also Section 3). In the final model, we hold the emission line
wavelength fixed at the rest wavelength, the width of \ion{C}{3}],
\ion{Si}{3}] and \ion{Al}{3} are constrained to be equal, and the
widths of the \ion{Fe}{3} 34 multiplet components are constrained to
be equal.

The spectrum from 1H~0707$-$495 appears to be more difficult to model
well.  There seems to be more \ion{Fe}{3} present in the spectrum of
this object.  A completely satisfactory fit could not be obtained.

\subsubsection{\ion{He}{2}}

This line was very difficult to identify and measure.  Following the
example of I~Zw~1, and the other high ionization lines, we expected it
to be strongly blueshifted and broad; i.e., from the wind.  Indeed, we
see no evidence for a narrow emission line at 1640 \AA\/.  We defined
a local continuum from apparently line-free region, and
measured the line using the wind template developed from the
\ion{C}{4} profile, as was done above.  This measurement is not very
satisfactory as there are other emission features of the same
amplitude, probably originating in \ion{Fe}{2}, near the feature that
we identify as \ion{He}{2}.   

\subsubsection{\ion{Mg}{2} region}

The spectrum around \ion{Mg}{2} is contaminated by broad humps of
\ion{Fe}{2} emission.  This makes measurement of the flux and width of
\ion{Mg}{2} difficult.  Following several other authors (e.g., Corbin
\& Boroson 1996; Dietrich et al.\ 2002), we develop an \ion{Fe}{2}
template from the medium-resolution {\it HST} spectrum from I~Zw~1
following the procedure used by Vestergaard \& Wilkes (2001).    We
adjust the amplitude of the \ion{Fe}{2} template so that it fits the
\ion{Fe}{2} feature in the spectra.  We then subtract the \ion{Fe}{2}
component, and fit the \ion{Mg}{2} line using the {\it LINER}
package.   It turns out we cannot constrain the doublet ratio as was
done for I~Zw~1 (Laor et al. 1997), so we require the doublet
component fluxes to be equal.  We cannot distinguish between Gaussian
and Lorenzian line profiles; however, we note that these lines are
significantly narrower than \ion{C}{3}] and \ion{Si}{3}]. 

\subsubsection{Other low-ionization lines}

The fluxes of several weak low-ionization lines were required for
photoionization modeling of the disk component in Paper 2.  All of the
lines considered in this section appeared to be centered at the rest
wavelength, and symmetric, so we infer that they are entirely produced
in the disk.  They are sufficiently weak that we could only measure
their flux with any confidence; their widths appear to be consistent
with the other low-ionization lines.  The \ion{C}{2} lines at
1335\AA\/ and 2327\AA\/ were modeled using Gaussians, employing {\it
LINER} to do the fitting.  \ion{C}{2}$\,\lambda 1335$\AA\/ was
slightly blended with an unidentified line at 1345\AA\/ (also seen in
I~Zw~1; see Laor et al.\ 1997) so these two lines were modelled
simultaneously.  This procedure worked well for IRAS~13224$-$3809.
1H~0707$-$495 posed a more difficult case, as there were several
strong absorption lines cutting through the \ion{C}{2}$\,\lambda
1335$\AA\/ lines.  The \ion{C}{2}$\,\lambda 2327$\AA\/ was more
difficult to model in both cases. It appeared to be perched on a
broad, blended \ion{Fe}{2} feature, and blended with \ion{Fe}{3}.  We
were able to measure it with some confidence for IRAS~13224$-$3809,
but did not obtain a measurement for 1H~0707$-$495.  We also measured
the flux and equivalent width of \ion{N}{3}]$\lambda 1750$\AA\/. This
was difficult, as it is barely detectable in IRAS~13224$-$3809, so
this measurement should be considered to be nearly an upper limit.  It
was more clearly detected in 1H~0707$-$495, and we measure the flux
and equivalent width by defining the local continuum on both sides of
the feature and integrating over the small excess at 1750\AA\/.

\subsubsection{H$\beta$ region}

We also obtained groundbased optical spectra from these two
objects. Near H$\beta$, the spectra in Narrow-line Seyfert 1 galaxies
is frequently cluttered with multiplets of \ion{Fe}{2} that make it
difficult to measure the properties of H$\beta$ and [\ion{O}{3}].
This problem is overcome by subtraction of an \ion{Fe}{2}
template developed from a high signal-to-noise spectrum from the
prototype NLS1 I~Zw~1 (e.g., Boroson \& Green 1992).  The \ion{Fe}{2}
subtracted spectra are displayed in Fig.\ 5.  The H$\beta$ line was
fitted with a Lorenzian profile to obtain the width.

\subsubsection{Comparison with Composite Spectra}

In Table 2, along with the emission-line measurements from
IRAS~13224$+$3809 and 1H~0707$-$495, we list measurements of emission
lines from three composite spectra: the LBQS composite (Francis et
al.\ 1991); the radio-quiet composite from {\it HST} spectra (Zheng et
al.\ 1997); a composite spectrum from the FIRST Bright Quasar Survey
(Brotherton et al.\ 2001).  In all cases, the equivalent widths of the
prominent features in our spectra are the same or smaller than the
equivalent widths in the composites, except for \ion{N}{5}, which has
larger equivalent widths in our spectra. Zheng et al.\ attempt to
deconvolve \ion{N}{5} from Ly$\alpha$; in that case, Ly$\alpha$ in our
spectra are about one third, and \ion{N}{5} are nearly double the
equivalent width of that in the Zheng composite.  The whole 1400\AA\/
feature and 1900\AA\/ feature have about the same equivalent width as
the composite. Zheng et al.\ (1997) attempt a deconvolution of the
1900\AA\/ feature; our spectra have the same \ion{Al}{3} equivalent
widths, larger by a factor of two \ion{Si}{3}] equivalent widths, and
much smaller (by a factor of 3) \ion{C}{3}] equivalent widths than the
Zheng composite.  The largest differences are in \ion{C}{4} and
\ion{Mg}{2}, which have much smaller (one fifth to one half) the
equivalent width of the composite spectra.  These comparisons
underline the fact that the UV emission lines, excepting \ion{N}{5},
are relatively weak in NLS1s.

\section{Comparison with Other NLS1s}

The UV and optical properties of IRAS 13224$-$3809 and 1H~0707$-$495,
including their luminosity, their emission line profiles, and their
equivalent widths and ratios, are almost identical.  How do their
properties compare with those of other NLS1s?  This is an important
question for two reasons.  First, their X-ray properties, while again
virtually identical to one another, are markedly different from those
of other NLS1s.  In a study of the X-ray properties of NLS1s using
{\it ASCA} data, Leighly (1999b) found that there is a correlation
between the prominence or strength of the soft excess\footnote{The
soft excess (e.g., Turner \& Pounds 1989) is the generic term for the
excess of emission in the soft (0.1--2.0 keV) X-ray band over the
power law obtained in the 2--10 keV band.} and the amplitude of the
variability, with IRAS~13224$-$3809 and 1H~0707$-$495 displaying among
the strongest soft excesses and highest-amplitude variability in the
sample.  Also, these objects are two of the three that possess an
unusual spectral feature near 1~keV that has been interpreted as
highly blueshifted absorption feature (Leighly et al.\ 1997) or as
strong Fe-L absorption (Nicastro, Fiore \& Matt 1999).  We would like
to know if the differences between IRAS 13224$-$3809 and 1H~0707$-$495
and other NLS1s that are found in the X-ray band are also seen in the
UV band.  The second reason we are motivated to compare these two
objects with other NLS1s involves our interpretation of our analysis
of the UV spectra that is discussed in detail in Paper II.  

To address these questions, we retrieved the spectra from a
heterogeneous group of 14 NLS1s from the {\it HST} archive, and
analyzed them.  Analysis of properties of NLS1 UV spectra has been
previously presented by several groups.  However, reanalysis is
necessary because we wish to investigate both the emission line
profiles and their equivalent widths and ratios.  The most detailed
investigation to date was presented by Kuraszkiewicz et al.\ (2000).
Our analysis differs from theirs in that they confined their analysis
to the emission-line fluxes and ratios, while we are interested also
in the emission-line profiles, in particular that of \ion{C}{4}, as
well.  A comparison of the blueshift of \ion{C}{4} with the width of
H$\beta$ has been published by Marziani et al.\ (1996).  However, they
do not consider the fluxes and ratios of other important lines in the
spectrum.

The following objects, listed in order of right ascension, were
investigated: Mrk~335, WPVS~007, I~Zw~1, Ton~S180, RX~J0134.2$-$4258,
PG~1001$+$054, RE~1034$+$39, PG~1115$+$407, PG~1211$+$143,
PG~1404$+$226, PG~1402$+$261, Mrk~478, Mrk~493, and Ark~564.  These
objects represent a range of redshifts from 0.024 to 0.235, and span
two orders of magnitude in monochromatic luminosity at 2500\AA\/; the
luminosities of IRAS~13224$-$3809 and 1H~0707$-$495 are in the middle
of the sample.  Almost all of these spectra were obtained using the
FOS detector and therefore we reprocessed all of them, using the Post
Operational Archives (POA)
calibration\footnote{http://www.stecf.org/poa/pcrel/POA\_CALFOS.html}
when appropriate.  Ton~S180 was observed with the STIS detector, and
we retrieved those data from the archive using ``on-the-fly''
reprocessing.

We checked the wavelength calibration of each spectrum using the
Galactic absorption lines.  The spectra were then shifted up to $\sim
1.5$\AA\/ as necessary.  Our first estimate of the redshift was
obtained from the spectra directly, by fitting the \ion{Mg}{2} lines.
We assume that \ion{Mg}{2} lines are centered at the systemic
redshift.  This is justified by the similarity between \ion{Mg}{2} and
H$\beta$ emission in AGNs (McLure \& Jarvis 2002), and the fact that
all of these objects have narrow H$\beta$ and \ion{Mg}{2}.  These
redshifts were in good agreement with those listed in NED.  The
spectra were then shifted into the rest frame.  The Galactic E(B-V)
was obtained from NED, and the spectra were corrected using the CCM
reddening law (Cardelli, Clayton \& Mathis 1989).

We used a procedure similar to that employed for IRAS~13224$-$3809 and
1H~0707$-$495 as described in Section 3.1 to investigate the profile
of the \ion{C}{4} line.  We first identified and subtracted the local
continuum, and excluded portions of the profile modified by Galactic
absorption lines.  We require a smoothed profile to study the shape of
the line.  We used two different 
methods depending on the signal-to-noise ratio to obtain the smoothed
profile. When the signal-to-noise ratio was relatively high, we fit
the continuum-subtracted spectrum in the region of the emission line
with a spline function as described in Section 2.3.1. When the signal
to noise is lower, the multinode spline could not be constrained, so 
instead we fit the profile with several Gaussians or Lorenzians, and
used the model as our smoothed profile.  In a few objects (Ark 564,
WPVS 007, PG 1404+226), the line profiles were altered by the presence
of strong associated absorption lines.  For these objects, the line
profiles were reconstructed using multiple-component fits, but there
is no doubt that the result is uncertain, especially with regard to
the equivalent widths.  Therefore, these objects are denoted by a
different symbol in plots that include \ion{C}{4} information.  A
broad absorption line is observed in PG 1001+054 (Brandt, Laor \&
Wills 2000). However, this broad absorption line appears to be
sufficiently detached that it does not alter the \ion{C}{4} line
profile.

We then measured a number of properties from the resulting \ion{C}{4}
profile, including the flux and equivalent width.  We investigated
several possible ways to parameterize the shape and offset of the
\ion{C}{4} line.  We considered using the velocity offset (e.g.,
Marziani et al.\ 1996); however, this does not take into account the
fact that the lines are very asymmetric.  We looked into using
the moments of the line profile (mean, variance, skew and kurtosis).
We also investigated using the fraction of the line blueward of the
rest wavelength (here taken to be 1549.5 \AA\/, the average of the
doublet wavelengths).  We found that the mean wavelength is strongly
correlated with the fraction of the emission blueward of 1549.5 \AA\/,
and finally opted to use the latter as the parameter representing the
asymmetry and blueshift of the line.  We also obtained the kurtosis as
a measurement of the peakiness of the line: a line with positive
kurtosis is pointy, and a line with negative kurtosis has a flattened
profile.

Other high ionization lines were measured.  We obtained a rough
estimate of the flux and equivalent width of \ion{N}{5} by modeling it
with several Gaussians or Lorenzians.  In some objects, the lines were
narrow enough that \ion{N}{5} could be cleanly separated from
Ly$\alpha$, and the principal uncertainty in the measurements of the
flux and equivalent width is the placement of the continuum.  In other
objects, \ion{N}{5} overlapped Ly$\alpha$; we estimate the uncertainty
in the flux and equivalent width may be as high as 40\% for these
objects.  As discussed in Section 2.3.3 , the \ion{O}{4}] and
\ion{Si}{4} lines near 1400\AA\ were quite difficult to separate in
IRAS~13224$-$3809 and 1H~0707$-$495. Thus, for these other objects, we
measure the flux and equivalent width of the whole 1400\AA\/ feature
by simply integrating over it after defining a local continuum.
\ion{He}{2} was also measured.  In some objects, this line was blended
with \ion{O}{3}]$\lambda \lambda 1660.8, 1666.1$; in these cases, we
used a model consisting of several Gaussians to isolate the
contribution of \ion{He}{2}.  In several objects, a significant source
of uncertainty was determination of the continuum when the relatively
weak \ion{He}{2} line was blueshifted and broad.  Finally, we computed
ratios of the fluxes of the high-ionization lines with that of
\ion{C}{4}.

The feature near 1900\AA\/ is comprised primarily of the intermediate
ionization lines \ion{Al}{3}, \ion{Si}{3}], \ion{C}{3}], and
\ion{Fe}{3} UV34.  This was modeled in a way similar to that described
in Section 2.3.4.  As discussed in Vestergaard \& Wilkes (2001), the
\ion{Fe}{3} UV34 multiplet (1894.8\AA\/, 1914.1\AA\/, 1926.3\AA\/;
$^7S-^7P^\circ$) contributes significant uncertainty to line
measurements in this region. However, in all of the objects that we
looked at, the intermediate ionization lines appeared to be narrow,
symmetric, and centered at their rest wavelengths.  We therefore
assumed that all of the lines in the 1900\AA\/ feature have the same
width, and assumed that they are centered at their rest wavelength.
The presence of \ion{Fe}{3} UV34 generally could be ascertained by the
presence of the 1926\AA\/ line, which was discernible from the wings
of \ion{C}{3}] lines in these narrow-line objects.  The gf values
obtained from The Atomic Line
list\footnote{\url{http://www.pa.uky.edu/$\sim$peter/atomic/}}
indicate that in the optically-thin case, the line ratios
1926:1914:1895 should be 1:1.4:1.8.  However, radiative transfer is
likely to be important for these lines.  In some cases, the spectra
were fit well assuming that the three \ion{Fe}{3} lines have equal
intensity, as would be appropriate if the gas is very optically thick.
However, in several objects (e.g., Ark 564; Fig.\ 15), the 1914\AA\/
component was much stronger than the other two; this enhancement was
discussed previously by Vestergaard \& Wilkes (2000) for I~Zw~1,
although these authors do not give an explanation for it.  It turns
out that the upper-level energy for this transition is $82333.92\rm
cm^{-1}$, corresponding to 1214.6\AA\/, which is just 1.1\AA\/ from
Ly$\alpha$. Therefore, the 1914\AA\/ component of the \ion{Fe}{3} UV34
feature is plausibly selectively excited by Ly$\alpha$.
Interestingly, this seems to be almost the only line emitted under
broad-line region conditions that can be excited from the ground state
by Ly$\alpha$; many more lines, including \ion{O}{1}$\lambda 1302$ for
example, may be pumped by Ly$\beta$, and \ion{Fe}{2} may be pumped by
Ly$\alpha$ from low-lying metastable levels (e.g., Sigut \& Pradhan
2003).  Therefore, we also fit with a model in which the 1895 and 1926
\AA\/ components are constrained to have equal intensities, and the
1914 \AA\/ intensity is left free. In this case, the 1914 \AA\/
component ranges from 1.8 to typically 3 times the strength of the
other two.  The fluxes of \ion{Al}{3}, \ion{Si}{3}], and \ion{C}{3}]
lines, and the velocity width of \ion{C}{3}] were measured, and the
ratios of \ion{Al}{3} and \ion{Si}{3}] to \ion{C}{3}] were formed.
The 1900\AA\/ feature is complex; thus, there is significant
uncertainty in the relative intensities of the \ion{Si}{3}] and
\ion{C}{3}] lines; \ion{Al}{3} was separated sufficiently from the the
other two that uncertainty in this line was dominated by the continuum
placement.

\placefigure{fig15}

We obtained the equivalent widths and velocity widths of the
\ion{Mg}{2} line after subtracting the UV \ion{Fe}{2} as described in
Section 2.3.6.  A by-product of the \ion{Fe}{2} subtraction procedure
is a measure of the slope of the continuum $\alpha_u$, where
$F(\lambda) \propto \lambda^{\alpha_u}$, in the UV between about
2200\AA\/ and 3050\AA\/ (see Moore \& Leighly, in prep.\ for details).
We also measured the monochromatic luminosity at restframe 2500\AA\/
and 1400\AA\/.  Finally, we used generally nonsimultaneous X-ray data
to estimate $\alpha_{ox}$, the point-to-point power-law energy index between
2500\AA\/ and 2~keV in the rest frame.  If available, {\it ASCA} data
were used because flux at 2~keV could be estimated better than from
{\it ROSAT} data in which 2~keV falls at the end of the bandpass.
Uncertainties in intrinsic $\alpha_{ox}$ originate in the lack of
simultaneity of the observations, and reddening or absorption arising
in the host galaxy or AGN.

\subsection{Correlations and Principle Components Analysis}

We performed a correlation analysis of the parameters discussed above,
using the Spearman rank correlation coefficient.  In a few cases,
parameters could not be measured: an example is RX~J0134.2$-$4258, in
which no \ion{C}{3}], \ion{Si}{3}] and \ion{Al}{3} were
detected.   We also do not use the velocity width of \ion{C}{3}]
for 1H~0707$-$495 because it was much larger than the widths of other
similar lines in that object (see Section 2.3.4).  Since we are only
missing one object (three in one case) among 16, we do not use a
survival analysis.  The probabilities of accidental correlation are
given in the correlation matrix in Table 3.

We performed a principle components analysis using the correlation
matrix.  Although the number of objects and parameters is small, we
obtain a fairly significant clustering of correlations.  The first three
eigenvectors are given in Table 4; they represent 78\% of the variance
in the sample.  We caution that our sample is small, so these
eigenvectors may not survive if faced with a larger sample.
Nevertheless, they provide a convenient organizational framework in
which to discuss the correlations. 

\subsubsection{Eigenvector 1: the \ion{C}{4} and other high-ionization
  line properties} 

Eigenvector 1 is aligned along the mutually strong correlations
between \ion{C}{4} equivalent width and profile shape.  A number of
other parameters are strongly correlated or anticorrelated with the
asymmetry of the \ion{C}{4} line.  These include the high-ionization
properties of the equivalent width of \ion{He}{2} and the
\ion{N}{5}/\ion{C}{4} ratio, the intermediate-ionization line
properties of the \ion{Al}{3}/\ion{C}{3}] ratio and the
\ion{Si}{3}]/\ion{C}{3}] ratio, and the continuum property
$\alpha_{ox}$.  A number of these are shown in Fig.\ 7.  Another
interesting correlation is the one  between the \ion{N}{5}/\ion{C}{4}
ratio and the \ion{Al}{3}/\ion{C}{3} ratio; we defer that discussion
until Section 3.1.3.  

\placefigure{fig7}

The \ion{C}{4} asymmetry parameter is strongly anticorrelated with the
\ion{C}{4} equivalent width; that is, the strongly blueshifted lines
have the lowest equivalent width.  This was previously reported in
Leighly 2001 for this sample, and has since then been recovered from
the SDSS quasars (Richards et al.\ 2002), and a similar correlation
was previously noted by Wills et al.\ (1993).  The kurtosis is
strongly anticorrelated with the asymmetry parameter, implying that
the more strongly blueshifted lines have a flattened rather than
peaked profile.  This result suggests the hypothesis that the narrow,
symmetrical part of \ion{C}{4} is an additional component that
increases the equivalent width of the whole line when present, as well
as reducing the cumulative offset from the rest wavelength.

To test this idea, we constructed a composite line consisting of the
wind profile from IRAS~13224$-$3809, and a narrow and symmetric line
that has a root-secant profile with a FWHM of $2000 \,\rm km\,s^{-1}$
and varying normalization. This FWHM was chosen because with it we
could best match the relationship between kurtosis and asymmetry
parameter (dashed line in the second panel of Fig.\ 7) when used in
combination with the IRAS~13224$-$3809 wind profile.  We assumed that
the wind component has a fixed equivalent width, and that the narrow
symmetric component contributes with strength varying from zero in the
case of IRAS~13224$-$3809.  We then measured the asymmetry parameter
for these simulated line profiles, and infer an equivalent width by
scaling the flux with that of IRAS~13224$-$3809.  This relation is
shown by the dashed line in the top panel of Fig.\ 7. Surprisingly,
despite the very simple assumptions, the correspondence is quite good.
We note that at least one other factor should influence the line
equivalent width: some objects have bluer continua than others.  In
fact, there is an anticorrelation between $\alpha_u$ and the asymmetry
parameter, such that objects with asymmetrical lines have bluer
continua, and therefore would have a lower equivalent width for the
same \ion{C}{4} line flux.  We note that this is not a clear
manifestation of the anticorrelation between line equivalent width and
luminosity (the Baldwin effect; Baldwin 1977);  Fig.\ 8 shows that
the objects with the lowest equivalent widths have mid-range
luminosities.  We discuss the Baldwin effect further in Section 3.4.3

The asymmetry parameter is anticorrelated with both continuum
parameters $\alpha_{ox}$ and $\alpha_u$.  Thus, objects with blueward
asymmetric lines have relatively bluer UV continua and relatively
lower X-ray flux.  An exception is I~Zw~1, which appears to have a
reddened UV continuum (Smith et al.\ 1997). This dependence can be
directly interpreted in terms of resonance line-driven winds in
AGN. In such winds, the acceleration comes from resonance scattering
of the UV continuum from the central engine or accretion disk by ions
in the wind.  An intrinsically blue continuum provides more photons to
be scattered; thus a blue continuum should be associated with an
asymmetric line profile, as is seen.  Both Murray et al.\ (1995) and
Proga, Stone \& Kallman (2000) discuss the fact that strong soft X-ray
emission tends to overionize the ions required for resonance
scattering; without these ions, the wind is suppressed.  Thus, lower
X-ray flux should be associated with the presence of the asymmetrical
line profile, as is observed.

We note that very few of the UV and X-ray observations were
coordinated, so relative variability between UV and X-ray observations
adds noise to $\alpha_{ox}$; however, such noise would tend to
decorrelate the parameters, rather than cause a false correlation.
Also, a few objects may be reddened (e.g., Ark~564; Crenshaw et al.\
2002); this would artificially increase $\alpha_{ox}$.  Also, PG
1001$+$054 has a broad absorption line (Brandt, Laor \& Wills 2000),
and a very steep $\alpha_{ox}$, which may imply that the X-rays are
absorbed.  Nevertheless, when the suspect objects are removed, the
trend remains ($P=1.9$\%), tentatively supporting our hypothesis.
Furthermore, we note that our hypothesis is supported by the
emission-line properties of two objects for which we do have
simultaneous UV and X-ray observations (RE~1034$-$39: Casebeer \&
Leighly 2004; PHL 1811: Leighly et al.\ in prep.).

Also associated with the asymmetry parameter and \ion{C}{4} equivalent
width is the equivalent width of \ion{He}{2}; objects with blueshifted
\ion{C}{4} have weak \ion{He}{2} lines.    This result has two
interpretations. It may be a consequence of the fact that in any
particular object, if one line has a high equivalent width, many of
the other lines also have a high equivalent widths, possibly a result
of a deficient continuum or large covering fraction, for example.
However, there is another interpretation for the anticorrelation
between the asymmetry parameter and the \ion{He}{2} equivalent width.
A weak \ion{He}{2} line is generally considered to be smoking-gun
evidence for a continuum weak in soft X-rays; if this is the factor
driving the correlation, it would also support the presence of
a resonance-line driven wind, as mentioned above.   

Interestingly, there is no similar correlation with the equivalent
width of the stronger high-ionization line, \ion{N}{5}.  There is an
weak correlation between \ion{N}{5} equivalent width and $\alpha_u$,
such that objects with lower \ion{N}{5} equivalent widths have bluer
spectra.  This could be because bluer continua yield lower equivalent
widths for the same line flux.  Stronger correlations are seen with
the ratio of \ion{N}{5} to \ion{C}{4}.  Some of these may be related
to \ion{C}{4} correlations, but others are more strongly correlated
with the ratio than with the \ion{C}{4} alone. The anticorrelation
between \ion{He}{2} and the ratio of \ion{N}{5} to \ion{C}{4} may
support the idea that objects having a wind may have \ion{N}{5}
selectively excited by Ly$\alpha$; this issue is discussed further in
Paper II.  Several correlations are found between this ratio and
properties of the intermediate-ionization lines; these will be
discussed in Section 3.1.3.

\subsubsection{Luminosity/Equivalent Width Correlations}

We found a strong anticorrelation between the \ion{C}{4} equivalent
width and the profile asymmetry.  It has been known for years that
there exists generally an anticorrelation between AGN emission line
equivalent widths and luminosity; this is known as the Balwin effect
(Baldwin 1977).  The physical origin of the Baldwin effect has
remained a puzzle for many years; however, recent work suggests that
it is a result of UV/X-ray continuum becoming softer as the luminosity
increases (Dietrich et al.\ 2002).  In this sample, the correlations
reported in Table 3 between line equivalent width and the luminosity
at 2500\AA\/ are weak; it may be that the correlations are not
stronger because we probe only a limited range of luminosity.

In Fig.\ 8 we compare the relationship between $\lambda L_\lambda$ and
equivalent width for our sample with the regression obtained for a
much larger sample by Dietrich et al.\ (2002). We note that for the
purpose of this plot, the luminosities were calculated assuming the
cosmology used by Dietrich et al.\ (2002): $H_0=65\rm
km\,s^{-1}\,Mpc$, $\Omega_M=0.3$, and $\Omega_\Lambda=0.7$. The most
remarkable feature of this plot is that, although there is still an
anticorrelation between equivalent width and luminosity for NLS1s, the
relationship for many of the emission lines is offset from that of the
average quasar; for a given luminosity, NLS1s have lower
equivalent-width emission lines. This peculiar property of NLS1s was
previously reported by Wilkes et al.\ (1999).    Interestingly,
\ion{N}{5} is the exception to the trend; the \ion{N}{5} equivalent
widths from NLS1s are consistent with those from the average quasar.  

\placefigure{fig8}

1H~0707$-$495 and IRAS~13224$-$3809 have the smallest \ion{C}{4}
equivalent widths in the sample; however, they have median luminosity,
so their small equivalent width is not a consequence of the Baldwin
effect.  Interestingly, for some of the lines, the equivalent widths
from these two objects is consistent with those of the other NLS1s;
for other lines, they are markedly lower.  

\subsubsection{Intermediate-ionization Line Properties} 

Eigenvector 2 is aligned along the equivalent widths of the
\ion{Si}{3}] and \ion{Al}{3}; the equivalent widths of these lines are
  strongly correlated.  Mutual correlations are found between the
  ratio of \ion{C}{3}] to
\ion{C}{4}, the equivalent widths of \ion{Al}{3} and
\ion{Si}{3}], and the ratios of \ion{Si}{3}] and \ion{Al}{3} to
\ion{C}{3}].  These interdependencies are illustrated in Fig.\ 9.  

\placefigure{fig9}

What is the origin of these interdependencies?  We discuss this
qualitatively here, and provide a more quantitative discussion,
including photoionization modeling, in Paper II.   First, it is possible
that the trends reflect a difference in density of the line-emitting 
region.   Because of its low critical density of $3.4 \times
10^9\rm\,cm^{-3}$, \ion{C}{3}] is expected to decrease with respect to
\ion{Si}{3}] (critical density of $1.04\times 10^{11} \rm\,cm^{-3}$)
and \ion{Al}{3} (a permitted line) as the density increases.  Then, as
density increases, \ion{Si}{3}] and \ion{Al}{3} are expected to take
over the cooling, resulting in an increase in both their equivalent
widths and their ratios with respect to \ion{C}{3}].  However, an
increase in density would predict a decrease in the
\ion{C}{3}]/\ion{C}{4} ratio, since the critical density of \ion{C}{4}
is $2.06 \times 10^{15}\rm\,cm^{-3}$. This would predict an
anticorrelation between \ion{C}{3}]/\ion{C}{4} and
\ion{Si}{3}]/\ion{C}{3}], rather than the correlation that we observe.

Another possible origin of these trends is a variation in the
ionization parameter.  A decrease in ionization parameter predicts an
increase in the ratio of \ion{C}{3}] to \ion{C}{4}.  It may
conceivably produce larger equivalent widths of \ion{Si}{3}] and
\ion{Al}{3}, because their lower ionization potential allows them to
be produced under lower ionization conditions, leading to larger
ratios of \ion{Si}{3}]/\ion{C}{3}] and \ion{Al}{3}/\ion{C}{3}].   

There is another possible explanation that we discuss in much greater
detail in Paper II; photoionization modeling is presented there that
supports this explanation.  If we assume that the objects that have
strongly blueshifted \ion{C}{4} lines are characterized by the
presence of a wind, it is possible that the continuum passes through
the wind, or is ``filtered'' by the wind, before it illuminates the
intermediate-ionization line-emitting region\footnote{We differentiate
between a ``shielded'' continuum, which is assumed to have been
transmitted through highly ionized gas (e.g., Murray et al.\ 1995),
and a ``filtered'' continuum, which is assumed to have been
transmitted through the wind while ionizing and exciting it before
illuminating the disk and producing the observed intermediate- and
low-ionization lines.  This point is expanded upon in Paper II.}.
After passing through the wind, the continuum will be somewhat weaker
in the hydrogen continuum and will have very few photons in the 
helium continuum.  Such a continuum is effectively softer overall, and
therefore lower-ionization species should dominate the cooling.  As
the continuum becomes softer, we see a decrease in the ratio of
\ion{C}{3}] to \ion{C}{4}, because the continuum is no longer hard
enough to ionize C+2 efficiently.  \ion{Al}{3} and \ion{Si}{3}] then
increase relative to \ion{C}{3}] because Al+2 and Si+2 have lower
ionization potentials than C+2.

Some of the intermediate-ionization emission line properties are
correlated with properties of the high-ionization lines.  There are
strong correlations between  \ion{Al}{3}/\ion{C}{3}] and
\ion{Si}{3}/\ion{C}{3}] ratios and the asymmetry parameter  (Fig.\
7).  This perhaps also supports the concept of filtering outlined
above, since that can be interpreted as a shift of the cooling in the
intermediate-line emitting region to ions with lower ionization
potentials when a wind is present.  We also see a correlation between
the intermediate-ionization line ratios and the  \ion{N}{5}/\ion{C}{4}
ratio that is particularly strong for the \ion{Al}{3} ratio.  Because
\ion{N}{5} overlaps Ly$\alpha$ in the windy objects, it is possible
that \ion{N}{5} is selectively excited by Ly$\alpha$ in these
objects.  Thus, this correlation could be interpreted as additional
support for the filtering scenario outlined above.  

\subsubsection{The Line Velocity Widths} 

The velocity widths appear in the third eigenvector.  The correlations
between H$\beta$, \ion{Mg}{2} and \ion{C}{3}] velocity widths are not
as strong as might be expected, as seen in Fig.\ 10.  The scatter
possibly originates in nonuniform measurement of the H$\beta$ widths;
many of these values were taken from the literature.  Also, the width
of \ion{C}{3}] is difficult to measure, as it is blended with
\ion{Si}{3}] and \ion{Fe}{3} UV 34.  We do observe a correlation
between the velocity width of \ion{Mg}{2} and the luminosity at
2500\AA\/.  This correlation is expected, as objects with higher
luminosities should have larger black holes and correspondingly BLR
velocities (Laor 1998); we discuss this correlation in a larger sample
in Leighly \& Moore 2004 (in prep.).  We also see anticorrelations
between the \ion{C}{3}]/\ion{C}{4} ratio and H$\beta$ velocity width,
and the \ion{He}{2}/\ion{C}{4} ratio and \ion{Mg}{2} velocity width.
We have no ready explanation for these correlations.

\placefigure{fig10}

\section{Summary}

\begin{itemize}
\item We present a detailed analysis of the UV spectra from two
Narrow-line Seyfert 1 galaxies, IRAS~13224$-$3809 and 1H~0707$-$495.
We find that their continua are as blue as that of the
average quasar.  We observe that the high-ionization emission lines
(including \ion{N}{5} and \ion{C}{4}) are broad (FWHM$\approx\rm
5000\rm \,km\,s^{-1}$), and strongly blueshifted, peaking at around
$2500\rm \,km\,s^{-1}$ and extending up to almost $\sim 10,000\rm \,
km\,s^{-1}$.  In contrast, the intermediate- and low-ionization lines
(e.g., \ion{C}{3}] and \ion{Mg}{2}) are narrow (FWHM
1000--1900$\rm\,km\,s^{-1}$) and centered at the rest wavelength.
\ion{Si}{3}] is prominent, and other low ionization lines (e.g.,
\ion{Fe}{2} and \ion{Si}{2}) are strong.  Based on these observations,
the working model that we adopt considers that the blueshifted
high-ionization lines come from a wind that is moving toward us, with
the receding side blocked by the optically thick accretion disk, and
the intermediate- and low-ionization lines are emitted in the
accretion disk atmosphere or low-velocity base of the wind.

\item The strongly blueshifted \ion{C}{4} profile suggests that it is
dominated by emission in the wind.  We used the \ion{C}{4} profile
to develop a template for the wind.  We then used this template,
plus a narrow and symmetric component representing the disk emission
to model the other bright emission lines. We inferred that the
high-ionization lines \ion{N}{5} and \ion{He}{2} are also dominated
by wind emission, and a part of Ly$\alpha$ is emitted in the wind.
The intermediate- and low-ionization lines \ion{Al}{3},
\ion{Si}{3}], \ion{C}{3}], and \ion{Mg}{2} are dominated by disk
emission, and a part of Ly$\alpha$ is disk emission as well.  The
1400\AA\/ feature, comprised of \ion{O}{4}] and \ion{Si}{4} was
difficult to model; however, it appears to include both disk and
wind emission. 

\item IRAS~13224$-$3809 and 1H~0707$-$495 have distinctive X-ray
properties among NLS1s; in order to determine whether their
distinctive properties carry over to the UV, we analyzed {\it HST}
archival spectra from 14 other NLS1s with a range of two orders of
magnitude in UV luminosity.  We find indeed that these two objects are
extreme in this sample in the following properties: strongly
blueshifted \ion{C}{4} line, the low equivalent widths of many of the
lines, particularly \ion{C}{4} and \ion{He}{2}, high
\ion{C}{3}]/\ion{C}{4}, \ion{Si}{3}]/\ion{C}{3}],
\ion{Al}{3}/\ion{C}{3}], and \ion{N}{5}/\ion{C}{4} ratios, steep
$\alpha_{ox}$, and blue UV continuum.  Correlation analysis finds a
number of strong correlations.  An anticorrelation between \ion{C}{4}
asymmetry and equivalent width suggests that the line is generally
composed of a highly asymmetric wind component and a narrower
symmetric component.  The anticorrelation between \ion{C}{4} asymmetry
and $\alpha_{ox}$ and $\alpha_{u}$, and with \ion{He}{2} suggests that
UV-strong and X-ray weak continua may be associated with a wind, as
would be expected if the acceleration mechanism is radiative-line
driving.  The dominance of \ion{Al}{3} and \ion{Si}{3}] over
\ion{C}{3}] suggests possibly that the continuum is transmitted
through the wind before it illuminates the intermediate-ionization line
emitting gas.

\item Paper II investigates the physical conditions in the
line-emitting gas through {\it Cloudy} modeling of the disk and wind
emission lines.  The results are used, in combination with a toy
dynamical model for the wind, to estimate the distance of the wind
from the continuum emitting region.  Further discussion of the results
is given, including a comparison with previous results.
\end{itemize}



\acknowledgments
KML thanks many people for useful discussions, including Fabrizio
Nicastro, and especially Jules Halpern and the OU AGN group (Darrin
Casebeer, Chiho Matsumoto, \& Larry Maddox).  We thank Thaisa
Storchi-Bergmann for obtaining and Mike Eracleous for reducing the
optical spectrum from 1H~0707$-$495. Support for proposal \# 7360 was
provided by NASA through a grant from the Space Telescope Science
Institute, which is operated by the Association of Universities for
Research in Astronomy, Inc., under NASA contract NAS 5-26555. This
research has made use of the NASA/IPAC Extragalactic Database (NED)
which is operated by the Jet Propulsion Laboratory, California
Institute of Technology, under contract with the National Aeronautics
and Space Administration. This research has made use of data obtained
from the High Energy Astrophysics Science Archive Research Center
(HEASARC), provided by NASA's Goddard Space Flight Center. KML and JRM
gratefully acknowledge support by NASA grant NAG5-10171 (LTSA).

\clearpage







\clearpage

\clearpage

\clearpage

\clearpage

\clearpage

\clearpage

\begin{figure}
\epsscale{1.0}
\plotone{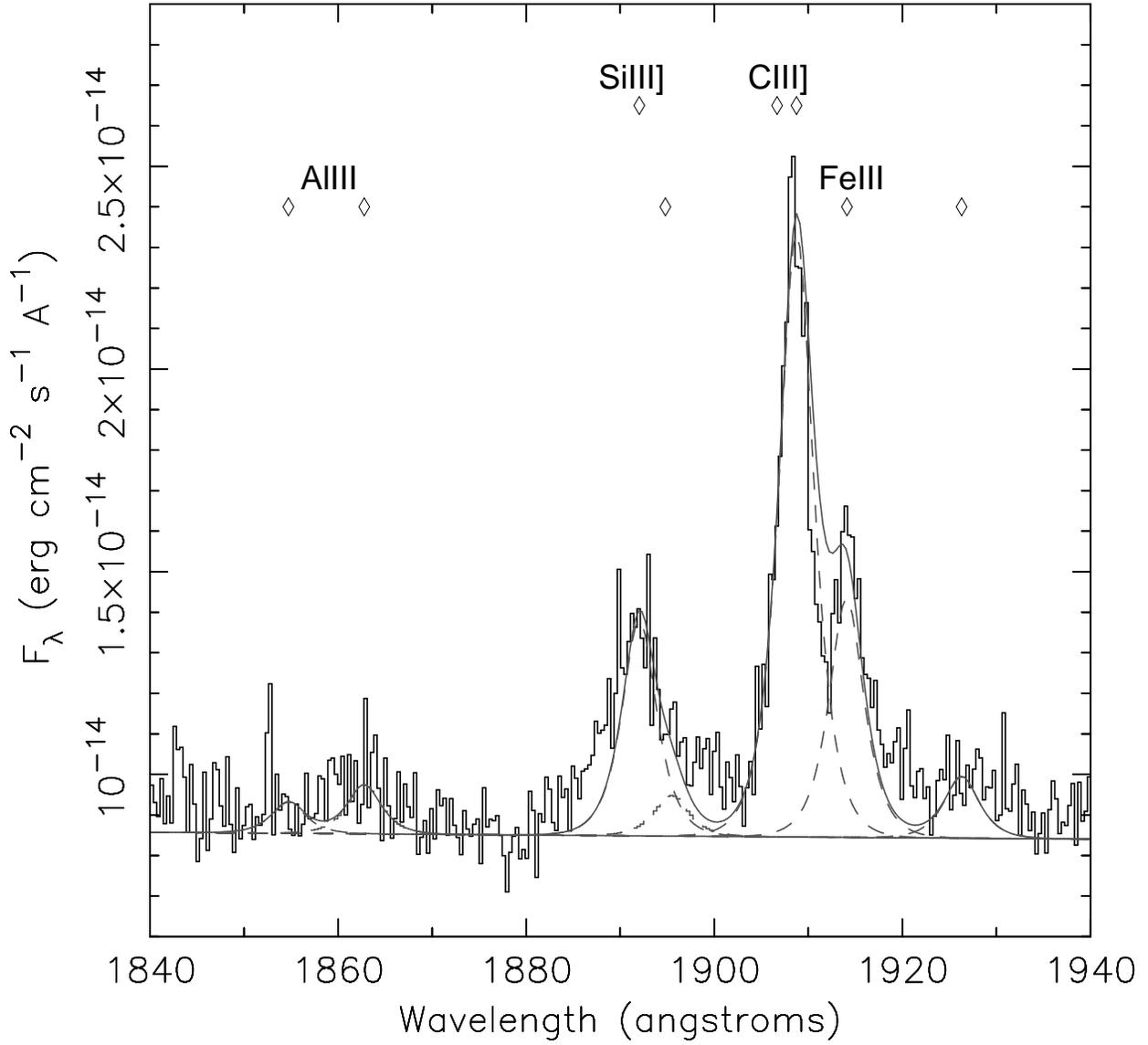}
\caption{Model for the 1900\AA\/ feature in Ark~564. Diamonds mark the
rest wavelength positions of the emission lines.  The very
narrow  lines allow the  pumped \ion{Fe}{3}$\lambda 1914$ feature to
be unambiguously detected.\label{fig6}} 
\end{figure}

\clearpage

\begin{figure}
\epsscale{0.25}
\plotone{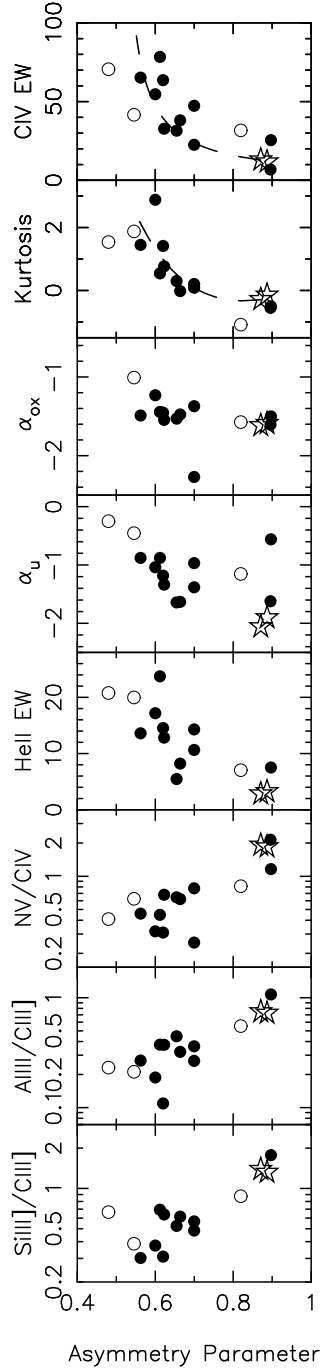}
\caption{Various parameters as a function of the asymmetry parameter.
Stars mark IRAS~13224$-$3809 and 1H~0707$-$495, and open circles mark
objects with significant \ion{C}{4} associated absorption lines;
their profiles and equivalent widths are uncertain because of
assumptions made to reconstruct the profiles (see Section 3).
Equivalent widths (EW) are given in units of Angstroms.  $\alpha_{ox}$
($F(E) \propto E^{\alpha_{ox}}$) 
is defined between 2500\AA\ and 2 keV in the rest frame.  $\alpha_{u}$
($F(\lambda) \propto \lambda^{\alpha_u}$) is obtained as a byproduct of
the UV \ion{Fe}{2} subtraction, and is defined between approximately
2200 \AA\/ and 3050 \AA\/.  Top and second panels: the
dashed line shows the predicted relationship between equivalent width
and kurtosis and asymmetry parameter for a two-component line
comprised of a wind and symmetric component (see text).\label{fig7}}
\end{figure}

\clearpage

\begin{figure}
\epsscale{0.3}
\plotone{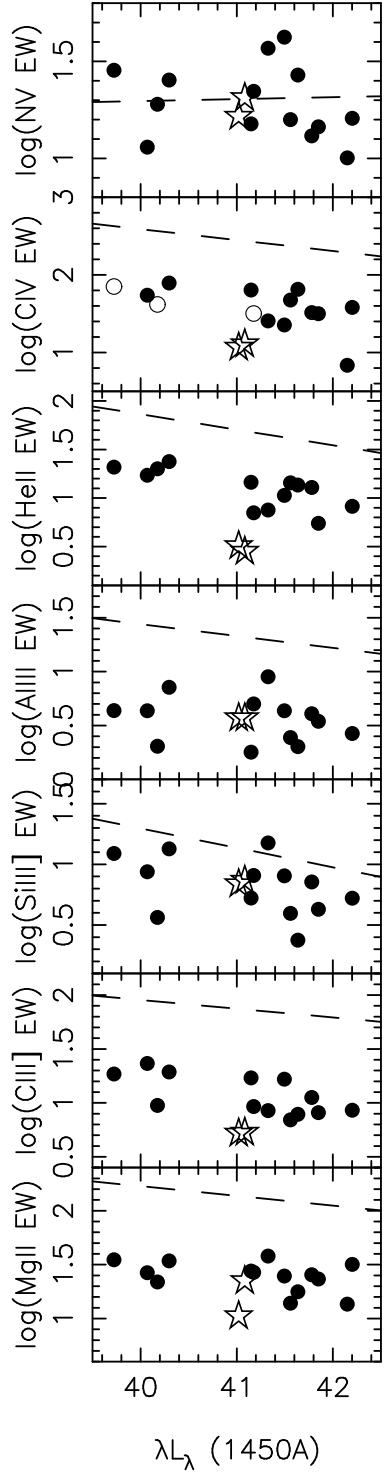}
\caption{Measured equivalent widths as a function of $\lambda
  L_\lambda$ at 1450\AA\/.  Symbols have the same meaning as in Fig.\ 7.
  Dashed lines are the Baldwin effect regressions computed by Dietrich
  et al.\ 2002.  Note that for the purpose of this plot, luminosities
  were computed using the cosmological parameters assumed by Dietrich et
  al.\ (2002).\label{fig8}}
\end{figure}

\clearpage

\begin{figure}
\epsscale{0.6}
\plotone{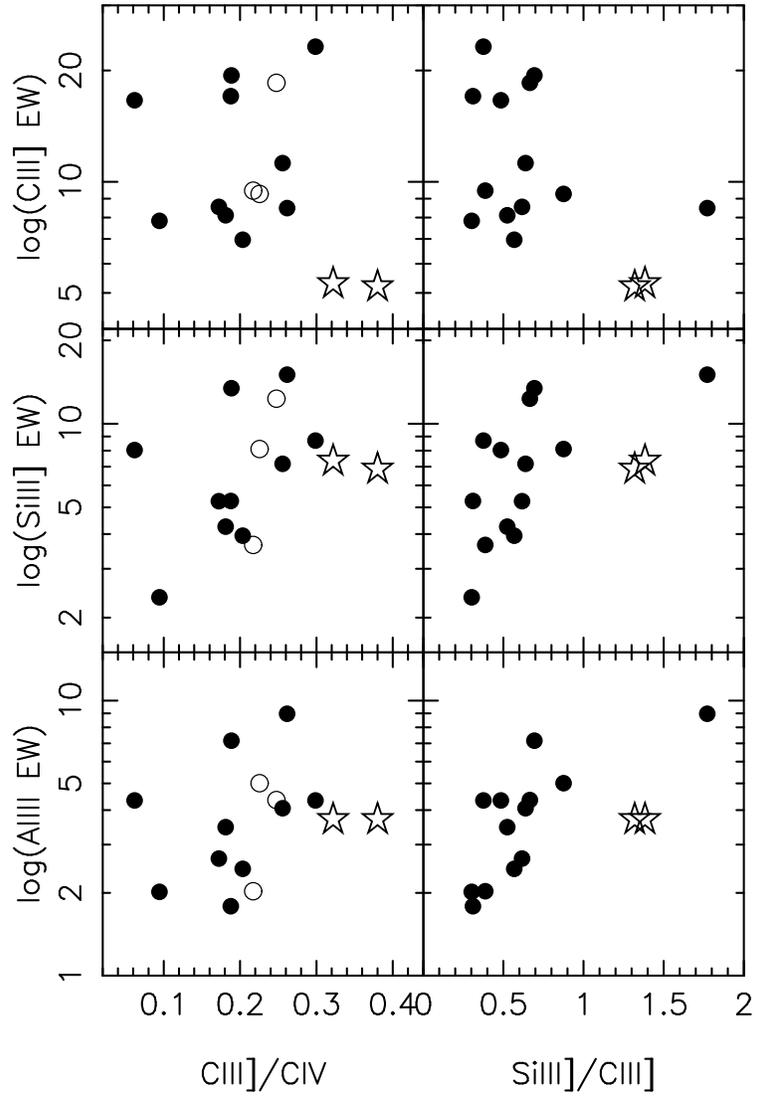}
\caption{Intermediate-ionization line equivalent widths as a function
  of the ratio of \ion{C}{3}] to \ion{C}{4} and the ratio of
  \ion{Si}{3}] to \ion{C}{3}]. Symbols have the same meaning as in
  Fig.\ 7. \label{fig9}}
\end{figure}

\clearpage

\begin{figure}
\epsscale{0.4}
\plotone{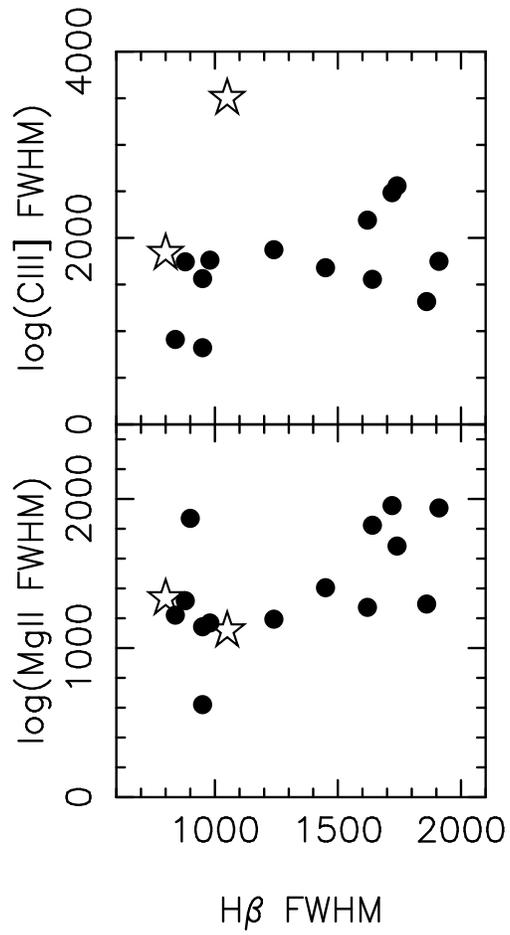}
\caption{Velocity widths of \ion{C}{2}] and \ion{Mg}{2} as a
  function of H$\beta$ velocity width.  Stars denote IRAS~13224$-$3809
  and 1H~0707$-$495.\label{fig10}} 
\end{figure}

\clearpage

\begin{deluxetable}{lcccccc}
\tabletypesize{\scriptsize}
\tablewidth{0pt}
\tablecaption{Observing log}
\tablehead{
\colhead{Target} & Date & \colhead{Spectrometer} & \colhead{Observed
  Wavelength} & \colhead{Resolution} & \colhead{Aperture\tablenotemark{a}} & \colhead{Exposure}  \\
&&& \colhead{(\AA\/)} & \colhead{(\AA\/)} & \colhead{(arc seconds)} &
\colhead{(seconds)}}
\startdata
IRAS 13224$-$3809 & 1999-06-05 & {\it HST} STIS G230L & 1568--3184 &
3.2 & $0.5 \times 0.24 $
& 4457 \\
& 1999-06-05 & {\it HST} STIS G140L & 1140--1730 & 1.2 & $0.5 \times 0.24$ & 8188 \\
&  1999-06-20 & CTIO 4m & 3870--7570 & 3.0 & $1 \times 4$ & 3600 \\
1H 0707$-$495 & 1999-02-16 & {\it HST} STIS G230L &  1568--3184 & 3.2
& $0.5 \times 0.24 $  & 2196 \\
& 1999-02-16  & {\it HST} STIS G140L & 1140--1730 & 1.2 & $0.5 \times 0.24$ &  5376 \\
& 1998-01-04 & CTIO 1.5m & 3500--6940 & 5.5 & $1.8 \times 7$ & 1800 \\
\enddata
\tablenotetext{a}{Slit width $\times$ extraction aperture in the
spatial direction}

\end{deluxetable}

\begin{deluxetable}{llllllllll}
\tabletypesize{\scriptsize}
\rotate
\tablewidth{0pt}
\tablecaption{Emission Line Measurements}
\tablehead{
\colhead{Emission line\tablenotemark{a}} & \multicolumn{3}{c}{IRAS 13224$-$3809}
& \multicolumn{3}{c}{1H 0707$-$495} & \multicolumn{3}{c}{Average EWs (\AA\/)}\\
 & \multicolumn{3}{c}{\hrulefill}
& \multicolumn{3}{c}{\hrulefill} &
\multicolumn{3}{c}{\hrulefill} \\
& \colhead{Flux} & \colhead{Equivalent Width} &
\colhead{Width} & 
\colhead{Flux} & \colhead{Equivalent Width} & \colhead{Width}
& \colhead{Francis} & \colhead{Zheng\tablenotemark{b}} & \colhead{Brotherton} \\
 & ($\times 10^{-14}\,\rm erg\,cm^{-2}\,s^{-1}$) & (\AA\/) & ($\rm
km\,s^{-1}$) &  ($\times 10^{-14}\,\rm erg\,cm^{-2}\,s^{-1}$) &
(\AA\/) & ($\rm km\,s^{-1}$) }
\startdata
Ly$\alpha$+\ion{N}{5} feature & 59.4 & 42 & & 166 & 45.9 & & 52 & & 87 \\
Ly$\alpha$ (broad) & 19.9 & 14.0 & & 65.3 & 17.7 & & & 85 \\
Ly$\alpha$ (narrow) & 8.3 & 6.0 & 1160 & 20.2 & 5.6 & 1135 \\
\ion{N}{5} (broad) & 27.9 & 20.5 && 57.8 & 16.4 & & & 10 \\
1400 \AA\ feature & 11.4 & 9.6 & & 28.3 & 9.9 & & 10 & 7.3 & 8 \\
\ion{Si}{4} (broad) & $5.5<7.2$\tablenotemark{c} & 
$4.6<6.1$\tablenotemark{c} & &  $10.1<15.9$\tablenotemark{c} & 
$3.5<5.6$\tablenotemark{c} \\
\ion{Si}{4} (narrow) & $0.75<3.7$\tablenotemark{c} & 
$0.63<3.2$\tablenotemark{c} & & $4.5<10.6$\tablenotemark{c} &
$1.6<3.7$\tablenotemark{c}  \\
\ion{O}{4} (broad) & $3.7<10.2$\tablenotemark{c} & 
$3.1<8.6$\tablenotemark{c} & & $9.4<22.4$\tablenotemark{c} 
& $3.3<7.9$\tablenotemark{c} \\
\ion{O}{4} (narrow) & $1.3<3.4$\tablenotemark{c} & 
$1.1<2.9$\tablenotemark{c} & & $2.3<7.6$\tablenotemark{c} 
& $0.8<2.7$\tablenotemark{c} \\
\ion{C}{4} (broad) & 14.6 & 12.8 & & 31.3 & 11.6 & & 37 & 59 & 33 \\
\ion{He}{2} (broad) & 3.1 & 2.8 & & 8.3 & 3.2 & & 12\tablenotemark{d} & 3.9 &
7.0\tablenotemark{d} \\
1900 \AA\ feature & 17.0 & 19.0 & & 42.5 & 18.3 & & 22 & & 17 \\
\ion{Al}{3} & 3.5 & 3.7 & 1835\tablenotemark{e} & 8.6 & 3.7 & 
3500\tablenotemark{e} & & 3.5 &    \\
\ion{Si}{3}] & 6.5 & 7.2 & 1835\tablenotemark{e}& 15.7 & 6.8 & 
3500\tablenotemark{e} & & 3.5 & \\
\ion{C}{3}] & 4.7 & 5.3 & 1835\tablenotemark{e} & 11.9 & 5.2 & 
3500\tablenotemark{e} & & 17.0  &\\
\ion{Mg}{2} &  6.7 & 17.3 & 866\tablenotemark{f} & 11.6 & 10.7 & 
1050\tablenotemark{f} & 50 & 64 & 34 \\
\ion{C}{2}$\lambda 1335$ & 3.5 & 2.9 & 1700 & 8.2 & 2.7 & 2100 & \\
\ion{C}{2}$\lambda 2327$ & 1.5 & 2.7 & & \\
\ion{N}{3}$\lambda 1750$ & 1.0 & 1.2 & & 4.4 & 2.0 \\
\enddata
\tablenotetext{a}{``Broad'' refers to lines fit with the \ion{C}{4}
template presumably coming from the wind, while ``narrow'' refers to
lines symmetric at zero velocity presumably coming from the disk or
low velocity base of the wind.}
\tablenotetext{b}{Radio-quiet quasars.}
\tablenotetext{c}{Best value and upper limit; see text for details.}
\tablenotetext{d}{\ion{He}{2}$\lambda 1640$+\ion{O}{3}]$\lambda 1663$}
\tablenotetext{e}{Widths constrained to be equal in fitting.}
\tablenotetext{f}{Width of one component of the doublet.}
\end{deluxetable}

\clearpage

\begin{deluxetable}{lcccccccccccc}
\tabletypesize{\scriptsize}
\rotate

\tablewidth{0pc}
\tablenum{3}
\tablecaption{Correlation Matrix}
\tablehead{
&  \multicolumn{3}{c}{\ion{C}{4} Properties} &
\multicolumn{5}{c}{High-ionization Line Properties} &
\multicolumn{3}{c}{Intermediate-ionization Line Properties} \\
& \multicolumn{3}{c}{\hrulefill} &
\multicolumn{5}{c}{\hrulefill} &
\multicolumn{3}{c}{\hrulefill} \\

\colhead{Property} &
\colhead{\ion{C}{4} EW} &
\colhead{Asymmetry} &
\colhead{Kurtosis} & 
\colhead{\ion{N}{5} EW} & 
\colhead{\ion{N}{5}/\ion{C}{4}} &
\colhead{\ion{He}{2} EW} & 
\colhead{\ion{He}{2}/\ion{C}{4}} &
\colhead{1400\AA\  EW} & 
\colhead{\ion{C}{3}] EW} & 
\colhead{\ion{C}{3}]/\ion{C}{4}} & 
\colhead{\ion{C}{3}] FWHM} &
 \\}

\startdata
\ion{C}{4} EW & $\cdots$ & $-5.2\times 10^{-3}$ & 0.10 & $\cdots$ &
$-$0.40 & $1.4 \times 10^{-3}$ & $\cdots$ & $\cdots$ & 2.1 & $\cdots$ & $-$4.7 \\

Asymmetry & $-5.2\times 10^{-3}$ & $\cdots$ & $-2.8\times 10^{-5}$ & $\cdots$ &
0.16 & $-$0.023 & $\cdots$ & $\cdots$ & $-$1.7 & $\cdots$ & 5.2   \\

Kurtosis & 0.10 & $-2.8\times 10^{-5}$ & $\cdots$ & $\cdots$ & $-$0.24 &
0.057 & $\cdots$ & $\cdots$ & 2.8 & $\cdots$ & $-$3.4 \\ 

\ion{N}{5} EW & $\cdots$ & $\cdots$ & $\cdots$ & $\cdots$ & $\cdots$ &
$\cdots$ & $\cdots$ & $\cdots$ &  $\cdots$ & $\cdots$ & $\cdots$ \\  

\ion{N}{5}/\ion{C}{4}  &  $-$0.40 & 0.16 & $-$0.24 & $\cdots$ & $\cdots$
& $-$0.51 & $\cdots$ & $\cdots$ & $-$0.069 & 2.2 & $\cdots$ \\ 

\ion{He}{2} EW & $1.4\times 10^{-3}$ & $-$0.023 & 0.057 & $\cdots$ &
$-$0.51 & $\cdots$ & $\cdots$ & 3.9 & 0.16 & $\cdots$ & $\cdots$  \\

\ion{He}{2}/\ion{C}{4} & $\cdots$ & $\cdots$ & $\cdots$ & $\cdots$ &
$\cdots$ & $\cdots$ & $\cdots$ & $\cdots$ & 
$\cdots$ & 2.0 & $\cdots$ \\

1400\AA\  EW & $\cdots$ & $\cdots$ & $\cdots$ & $\cdots$ & $\cdots$ &
3.9 & $\cdots$ & $\cdots$ & 
1.8 & $\cdots$ & $\cdots$ \\ 

\ion{C}{3}] EW & 2.1 & $-$1.7 & 2.8 & $\cdots$ &
  $-$0.069 & 0.16  & $\cdots$ & 1.8 & $\cdots$ & $\cdots$ & $\cdots$ \\  

\ion{C}{3}]/\ion{C}{4} & $\cdots$ & $\cdots$ & $\cdots$ & $\cdots$
  & 2.2 & $\cdots$ & 2.0 & $\cdots$ & $\cdots$ & $\cdots$ & $\cdots$ \\  

\ion{C}{3}] FWHM & $-$4.7 & 5.2 & $-$3.4 & $\cdots$ &  $\cdots$ &
  $\cdots$ & $\cdots$ & $\cdots$ &  $\cdots$ & $\cdots$ & $\cdots$ \\

\ion{Al}{3} EW &$\cdots$ & $\cdots$ & $\cdots$ & $\cdots$ & $\cdots$
& $\cdots$ & $\cdots$ & $\cdots$ & 
$\cdots$ & $\cdots$ & $\cdots$\\ 

\ion{Al}{3}/\ion{C}{3}] & $-$1.4 & 0.078 & $-$0.022 &   $\cdots$ &
  $8.0 \times 10^{-3}$ & $-$0.26 & $\cdots$ & $\cdots$ &  $-$1.3 &
  $\cdots$ & $\cdots$  \\

\ion{Si}{3}] EW & $\cdots$ & $\cdots$ & $\cdots$ & $\cdots$ &
  $\cdots$ & $\cdots$ & $\cdots$ & 2.3 &  $\cdots$ &
  $\cdots$ & $\cdots$ \\

\ion{Si}{3}]/\ion{C}{3}] & $\cdots$ & 0.86 & $-$0.22 & $\cdots$ &
    0.30 & $\cdots$ & $\cdots$ &  $\cdots$ & $\cdots$ & 1.7 & $\cdots$ \\ 

\ion{Mg}{2} EW & $\cdots$  & $\cdots$ & $\cdots$ & $\cdots$ &
$\cdots$ & $\cdots$ &  $\cdots$ & 2.5 & 1.1 & $\cdots$ &
$\cdots$ \\

\ion{Mg}{2} FWHM & $\cdots$ & $\cdots$ & $\cdots$ & $\cdots$ &
  $\cdots$ & $\cdots$ & $-$0.45 & $\cdots$ & 
$\cdots$ & $\cdots$ & $\cdots$  \\

H$\beta$ FWHM & $\cdots$ & $\cdots$ & $\cdots$ & $\cdots$ & $\cdots$
  & $\cdots$ & $-$4.2 & $\cdots$ & $\cdots$ & $-$0.29 & $\cdots$  \\

$\alpha_{ox}$ & 0.028 & $-$0.50 & 0.48 & $\cdots$ & $\cdots$
  & 0.043 & 4.7 & $\cdots$ & $\cdots$ & $\cdots$ & $-$0.87 \\  

$\alpha_{u}$ & 1.9 & $-$2.7 & 5.5 & 2.5 & $-$4.4 &  0.45 & $\cdots$ & 0.30 & 
3.1 & $\cdots$ & $\cdots$ \\

$L_{2500}$  & $\cdots$ & $\cdots$ & $\cdots$ & $\cdots$ & $\cdots$ &
$\cdots$ & $\cdots$ & $\cdots$ & $-$4.4 & $\cdots$  & $\cdots$ \\

\enddata
\tablecomments{Percentage probability that the observed correlation is
  accidental are listed when the value is less than 6 (i.e., greater
  than 94\% confidence that correlation is real).}

\end{deluxetable}

\clearpage

\begin{deluxetable}{lccccccccccc}
\tabletypesize{\scriptsize}
\rotate

\tablewidth{0pc}
\tablenum{3}
\tablecaption{Correlation Matrix (cont.)}
\tablehead{
&  \multicolumn{4}{c}{Intermediate-ionization Line Properties (cont.)}
& \multicolumn{3}{c}{Low-ionization Line Properties} &
\multicolumn{3}{c}{Continuum Properties} \\
 & \multicolumn{4}{c}{\hrulefill}
& \multicolumn{3}{c}{\hrulefill} &
\multicolumn{3}{c}{\hrulefill} \\

\colhead{Property} &
\colhead{\ion{Al}{3} EW} & 
\colhead{\ion{Al}{3}/\ion{C}{3}]} &
\colhead{\ion{Si}{3}] EW} & 
\colhead{\ion{Si}{3}]/\ion{C}{3}]} &
\colhead{\ion{Mg}{2} EW} & 
\colhead{\ion{Mg}{2} FWHM} &
\colhead{H$\beta$ FWHM} &
\colhead{$\alpha_{ox}$} & 
\colhead{$\alpha_{u}$} & 
\colhead{$L_{2500}$}  \\}

\startdata

\ion{C}{4} EW &$\cdots$ & $-$1.4 & $\cdots$ & $\cdots$ & $\cdots$ &
$\cdots$ & 
$\cdots$ & 0.028 & 1.9 & $\cdots$\\

Asymmetry & $\cdots$ & 0.078 & $\cdots$ & 0.86 & $\cdots$ &
$\cdots$ & 
$\cdots$ & $-$0.50 & $-$2.7 & $\cdots$ \\

Kurtosis & $\cdots$ & $-$0.022 & $\cdots$ & $-$0.22 & $\cdots$ &
$\cdots$ & 
$\cdots$ & 0.48 & 5.5 & $\cdots$ \\ 

\ion{N}{5} EW & $\cdots$ & $\cdots$ & $\cdots$ & $\cdots$ & $\cdots$ &
$\cdots$ & 
$\cdots$ & $\cdots$ & 2.5 &  $\cdots$\\  

\ion{N}{5}/\ion{C}{4}  & $\cdots$ & $8.0\times 10^{-3}$ & $\cdots$ & 0.30
& $\cdots$ & $\cdots$ & 
$\cdots$ & $\cdots$ & $-$4.4 & $\cdots$\\ 

\ion{He}{2} EW &  $\cdots$ & $-$0.26 & $\cdots$ & $\cdots$ &
$\cdots$ & $\cdots$ & 
$\cdots$ & 0.043 & 0.45 & $\cdots$ \\

\ion{He}{2}/\ion{C}{4} & $\cdots$ & $\cdots$ & $\cdots$ & $\cdots$ &
$\cdots$ & $-$0.45 & 
$-$4.2 & 4.7 & $\cdots$ & $\cdots$
\\

1400\AA\  EW & $\cdots$ & $\cdots$ & 2.3 & $\cdots$ &
2.5 & $\cdots$ & 
$\cdots$ & $\cdots$ & 0.30 & $\cdots$
\\ 

\ion{C}{3}] EW &$\cdots$ & $-$1.3 & $\cdots$ & $\cdots$ & 1.1
  & $\cdots$ & 
$\cdots$ & $\cdots$ & 3.1 & $\cdots$ \\  

\ion{C}{3}]/\ion{C}{4} &$\cdots$ & $\cdots$ & $\cdots$ & 1.7 &
  $\cdots$ & $\cdots$ & 
$-$0.29 & $\cdots$ & $\cdots$ & $-$4.4 \\  

\ion{C}{3}] FWHM & $\cdots$ & $\cdots$ & $\cdots$ & $\cdots$ &
  $\cdots$ & $\cdots$ & 
$\cdots$ & $-$0.87 & $\cdots$ & $\cdots$ \\

\ion{Al}{3} EW &  $\cdots$ & $\cdots$ & $1.3 \times 10^{-4}$ & 0.79 &
2.4 & $\cdots$ & 
$\cdots$ & $\cdots$ & $\cdots$ & $\cdots$ \\ 

\ion{Al}{3}/\ion{C}{3}] & $\cdots$ & $\cdots$ & $\cdots$ &
  0.010 & $\cdots$ & $\cdots$ & 
$\cdots$ & $-$4.3 & $\cdots$ & $\cdots$\\

\ion{Si}{3}] EW & $1.3 \times 10^{-4}$ & $\cdots$ & $\cdots$ & 2.4 &
  0.28 & $\cdots$ & 
$\cdots$ & $\cdots$ & $\cdots$ & $\cdots$  \\  

\ion{Si}{3}]/\ion{C}{3}]     & 0.79 & 0.010 & 2.4 &
    $\cdots$ & $\cdots$ & $\cdots$ & 
$\cdots$ & $\cdots$ & $\cdots$ & $\cdots$\\

\ion{Mg}{2} EW & 2.4 & $\cdots$ & 0.28 & $\cdots$ &
$\cdots$ & $\cdots$ & 
$\cdots$ & $\cdots$ & $\cdots$ & $\cdots$\\

\ion{Mg}{2} FWHM &  $\cdots$ & $\cdots$ & $\cdots$ & $\cdots$ &
  $\cdots$ & $\cdots$ & 
$\cdots$ & $\cdots$ & $\cdots$ & 0.40 \\

H$\beta$ FWHM & $\cdots$ & $\cdots$ & $\cdots$ & $\cdots$ & $\cdots$
  & $\cdots$ & 
$\cdots$ & $\cdots$ & $\cdots$ & 5.9 \\

$\alpha_{ox}$ & $\cdots$ & $-$4.3 & $\cdots$ & $\cdots$ & $\cdots$ &
  $\cdots$ & 
$\cdots$ & $\cdots$ & $\cdots$ & $\cdots$\\  

$\alpha_{u}$ & $\cdots$ & $\cdots$ & $\cdots$ & $\cdots$ & $\cdots$ &
  $\cdots$ & 
$\cdots$ & $\cdots$ & $\cdots$ & $\cdots$ \\ 

$L_{2500}$  &  $\cdots$ & $\cdots$ & $\cdots$ & $\cdots$ & $\cdots$ &
  0.40 & 
5.9 & $\cdots$ & $\cdots$ & $\cdots$  \\

\enddata
\tablecomments{Percentage probability that the observed correlation is
  accidental are listed when the value is less than 6 (i.e., greater
  than 94\% confidence that correlation is real).}

\end{deluxetable}

\setcounter{table}{3}
\normalsize

\begin{deluxetable}{lrrr}
\tablewidth{0pt}
\tablecaption{Principal Components Analysis Results}
\tablehead{
\colhead{Parameter} & \colhead{Eigenvector 1} & \colhead{Eigenvector
  2} & \colhead{Eigenvector 3} \\}
\startdata

\ion{C}{4} EW  &  0.32 &  0.02 &  0.01 \\
Asymmetry & $-$0.32 &  0.06 &  0.01 \\
Kurtosis &  0.32 & $-$0.08 & $-$0.09 \\
\ion{N}{5} EW &  0.02 &  0.21 &  0.27 \\
\ion{N}{5}/\ion{C}{4} & $-$0.29 &  0.05 & $-$0.24 \\
\ion{He}{2} EW &  0.33 &  0.07 & $-$0.05 \\
\ion{He}{2}/\ion{C}{4} &  0.06 &  0.15 & $-$0.42 \\
1400\AA\  EW   &  0.17 &  0.26 &  0.18 \\
\ion{C}{3}] EW   &  0.26 &  0.16 &  0.14 \\
\ion{C}{3}]/\ion{C}{4} & $-$0.11 &  0.27 & $-$0.33 \\
\ion{C}{3}] FWHM & $-$0.21 &  0.06 &  0.31 \\
 \ion{Al}{3} EW  & $-$0.06 &  0.40 &  0.16 \\
\ion{Al}{3}/\ion{C}{3}]  & $-$0.30 &  0.16 & $-$0.02 \\
\ion{Si}{3}] EW & $-$0.01 &  0.42 &  0.15 \\
\ion{Si}{3}]/\ion{C}{3}]  & $-$0.23 &  0.32 & $-$0.02 \\
\ion{Mg}{2} EW  &  0.12 &  0.26 &  0.27 \\
\ion{Mg}{2}] FWHM & $-$0.09 & $-$0.23 &  0.30 \\
H$\beta$] FWHM   &  0.06 & $-$0.23 &  0.34 \\
$\alpha_{ox}$  &  0.28 & $-$0.02 & $-$0.20 \\
$\alpha_{u}$  &  0.24 &  0.19 &  0.12 \\
$L_{2500}$   & $-$0.17 & $-$0.27 &  0.20 \\
\hline
Cumulative Fraction of Variance & 0.38 & 0.60 & 0.78 \\

\enddata

\end{deluxetable}




\begin{thebibliography}{}


\bibitem[Baldwin 1977]{baldwin77} Baldwin, J.\ A., 1977, ApJ, 214, 679
\bibitem[Baldwin 1997]{baldwin97} Baldwin, J.\ A., 1997, Proc.\
  ``Emission Lines in Active Galaxies: New Methods and Techniques'',
  eds.\ B.\ M.\ Peterson, F.-Z.\ Cheng, \& A.\ S.\ Wilson (ASP: San
  Francisco) p.\ 80
\bibitem[Baldwin et al.\ 1996]{bfkchpvww96} Baldwin, J.\ A., Ferland,
G.\ J., Korista, K.\ T., Carswell,  R.\ F., Hamann, F., Phillips, M.\
M., Verner, D., Wilkes, B.\ J., \& Williams, R.\ E., 1996, ApJ, 461,
664 
\bibitem[Ballantyne, Iwasawa \& Fabian 2001]{bif01} Ballantyne, D.\
  R., Iwasawa, K., \& Fabian, A.\ C., 2001, MNRAS, 323, 506
\bibitem[Blades et al. 1988]{blades88} Blades, J.\ C., Wheatley, J.\
  M., Panagia, N., Grewing, M., Pettini, M., \& Wamsteker, W., 1988,
  ApJ, 334, 309
\bibitem[Boller et al.\ 1993]{bol93} Boller, Th., Tr\"umper, J.,
  Molendi, S., Fink, H., Schaeidt, S., Caulet, A., \& Dennefeld, M.,
  1993, A\&A, 279, 53
\bibitem[Boller, Brandt \& Fink 1996]{bol96} Boller, Th., Brandt, W.\
  N., \& Fink, H., 1996, A\&A, 305, 53
\bibitem[Boller et al.\ 1997]{bol97} Boller, Th., Brandt, W.\ N.,
  Fabian, A.\ C., \& Fink, H.\ H., 1997, MNRAS, 289, 393
\bibitem[Boller, Brandt \& Fink 1996]{bbf96} Boller, T., Brandt, W.\
  N., \& Fink, H., 1996, A\&A, 305, 53
\bibitem[Boller et al.\ 2002]{bol02} Boller, Th., Fabian, A.\ C.,
  Sunyaev, R., Tr\"umper, J., Vaughan, S., Ballantyne, D.\ R., Brandt,
  W.\ N., Keil, R., \& Iwasawa, K., 2002, MNRAS, 329, L1
\bibitem[Boller et al.\ 2003]{bol03} Boller, Th., Tanaka, Y., Fabian,
  A.\ C., Brandt, W.\ N., Gallo, L., Anabuki, N., Haba, Y., \&
  Vaughan, S., 2003, MNRAS, in press
\bibitem[Boroson 2002]{b02} Boroson, T.\ A., 2002, ApJ, 565, 78
\bibitem[Boroson \& Green 1992]{bg92} Boroson, T.\ A., \& Green, R.\
  F., 1992, ApJS, 80, 109
\bibitem[Boroson \& Meyers 1992]{bm92} Boroson, T.\ A., \& Meyers, K.\
  A., 1992, ApJ, 397, 442
\bibitem[Brandt, Mathur \& Elvis 1997]{bme97} Brandt, W.\ N., Mathur,
  S., \& Elvis, M., 1997, MNRAS, 285, L25
\bibitem[Brandt, Laor \& Wills 2000]{blw00} Brandt, W.\ N., Laor, A.,
  Wills, B.\ J., 2000, ApJ, 528, 637
\bibitem[Brotherton et al.\ 2001]{btbglw01} Brotherton, M.\ S., Tran,
  H.\ D., Becker, R.\ H., Gregg, M.\ D., Laurent-Muehleisen, S.\ A.,
  \& White, R.\ L., 2001, ApJ, 546, 775
\bibitem[Cardelli, Clayton \& Mathis 1989]{ccm89} Cardelli, J.\ A.,
  Clayton, G.\ C., \& Mathis, J.\ S., 1992, ApJ, 345, 245
\bibitem[Casebeer and Leighly 2004]{CL04} Casebeer, D., \& Leighly,
K.\ M., 2004, to be submitted to ApJ
\bibitem[Collin-Souffrin et al.\ 1988]{collin88} Collin-Souffrin, S.,
  Dyson, J.\ E., McDowell, J.\ C., \& Perry, J.\ J., 1988, MNRAS, 232, 539
\bibitem[Comastri et al. 1998]{cfgms89} Comastri A., et al.\ 1998,
  A\&A, 333, 31
\bibitem[Corbin \& Boroson 1996]{cb96} Corbin, M.\ R., \& Boroson, T.\
  A., 1996, ApJS, 107, 69
\bibitem[Crenshaw et al.\ 2002]{cren02} Crenshaw, D.\ M., et al.,
  2002, ApJ, 566, 187
\bibitem[Crenshaw, Kraemer, \& Gabel 2003]{ckg03} Crenshaw, D.\ M.,
  Kraemer, S.\ B., \& Gabel, J.\ R., 2003, AJ, 126, 1690
\bibitem[Dewangen et al.\ 2002]{dbsl02} Dewangen, G.\ C., Boller, Th.,
  Singh, K.\ P., \& Leighly, K.\ M., 2002, A\&A, 390, 65
\bibitem[Dietrich et al.\ 2002]{dietrich02} Dietrich, M., Hamann, F.,
  Shields, J.\ C., Constantin, A., Vestergaard, M., Chaffee, F.,
  Foltz, C.\ B., \& Junkkarinen, V.\ T., 2002, ApJ, 581, 912
\bibitem[Ferland (2001)]{fer01} Ferland, G.\ J., 2001, ``Hazy, a Brief
  Introduction to Cloudy 96.00''
\bibitem[Flower and Nussbaumer (1975)]{flow75} Flower, D.\ R., \&
Nussbaumer, H., 1982, A\&A, 45, 145
\bibitem[Francis et al.\ 1991]{fhfcwm91} Francis, P.\ J., Hewett, P.\
C., Foltz, C.\ B., Chaffee, F.\ H., Weymann, R.\ J., \& Morris, S.\
L., 1991, ApJ, 373, 465
\bibitem[Goodrich 1989]{goodrich89} Goodrich, R.\ W.\ 1989, ApJ, 342, 224
\bibitem[Graham, Clowes \& Campusano 1996]{gcc96} Graham, M.\ J.,
Clowes, R.\ G., \& Campusano, L.\ E., 1996, \mnras, 279, 1349
\bibitem[Grupe et al.\ 1998]{grupe98} Grupe, D., Beuermann, K.,
  Thomas, H.-C., Mannheim, K., \& Fink, H.\ H., 1998, A\&A, 330, 25
\bibitem[Grupe \& Leighly 2002]{gl02} Grupe, D., \& Leighly, K.\ M.\,
  2002, Proc.\ ``Workshop on
  X-ray Spectroscopy of AGN with Chandra and XMM--Newton'', eds.\ Th.\
  Boller, S.\ Komossa, S.\ Kahn, H.\ Kunieda, \& L.\ Gallo (MPE:
  Garching) p.\ 259
\bibitem[Grupe et al. 2000]{gltl00} Grupe, D., Leighly, K.\ M.,
Thomas, H.-C., \& Laurent-Muehleisen, S.\ A., 2000, A\&A, 356, 11
\bibitem[Grupe et al. 2001]{gtl01} Grupe, D., Thomas, H.-C., \&
Leighly, K.\ M., 2001, A\&A, 369, 450
\bibitem[Harper et al.\ 1999]{harper99} Harper, G.\ M., Jordan, C.,
  Judge, P.\ G., Robinson, R.\ D., Carpenter, K.\ G., \& Brage, T.,
  1999, MNRAS, 303, 41
\bibitem[Hartig \& Baldwin 1986]{hb86} Hartig, G.\ F., \& Baldwin, J.\
A., 1986, \apj, 302, 64
\bibitem[Kuraszkiewicz et al. 2000]{kwcm00} Kuraszkiewicz, J., Wilkes,
B.\ J., Czerny, B., \& Mathur, S., 2000, ApJ, 542, 692
\bibitem[Laor 1998]{l98}Laor, A., 1998, ApJL, 505, 83
\bibitem[Laor et al.\ 1997]{ljgb97} Laor, A., Jannuzi, B.\ T., Green,
  R.\ F., \& Boroson, T.\ A., 1997b, ApJ, 489, 656
\bibitem[Leighly et al.\ 1997]{L97} Leighly, K.\ M., Mushotzky, R.\
  F., Nandra, K., \& Forster, K., 1997, ApJL, 489, 25
\bibitem[Leighly 1999a]{L99a} Leighly, K.\ M., 1999a, ApJS, 125, 297
\bibitem[Leighly 1999b]{L99b} Leighly, K.\ M., 1999b, ApJS, 125, 317
\bibitem[Leighly 2000]{L00} Leighly, K.\ M., 2000, NewAR, 44, 395
\bibitem[Leighly 2004a]{L04a} Leighly, K.\ M., 2004, submitted to ApJ
  (Paper II)
\bibitem[Leighly 2004b]{L04b} Leighly, K.\ M., 2004, Proc.\
  ``Stellar-Mass, Intermediate-Mass, and Supermassive Black Holes'',
  eds.\ K.\ Makishima \& S.\ Mineshige, in press
\bibitem[Leighly et al. 2001]{lhhbi01} Leighly, K.\ M., Halpern, J.\
P., Helfand, D.\ J., Becker, R.\ H., \& Impey, C.\ D., 2001, AJ, 121,
2889 
\bibitem[Leighly 2001]{L01} Leighly, K.\ M., 2001, in ``Probing the
  Physics of Active Galactic Nuclei'', Eds.\ B.\ M.\ Peterson, R.\ W.\
  Pogge, \& R.\ S.\ Polidan (ASP: San Francisco), p.\ 293
\bibitem[Leighly et al. 2002]{lzkm02} Leighly, K.\ M., Zdziarski, A.\
  A., Kawaguchi, T., \& Matsumoto, C., 2002, Proc.\ ``Workshop on
  X-ray Spectroscopy of AGN with Chandra and XMM--Newton'', eds.\ Th.\
  Boller, S.\ Komossa, S.\ Kahn, H.\ Kunieda, \& L.\ Gallo (MPE:
  Garching) p.\ 259
\bibitem[Marziani et al.\ 1996]{msdcm96} Marziani, P., Sulentic, J.\
  W., Dultzin-Hacyan, D., Calvani, M., \& Moles, M., 1996, ApJS, 104, 37
\bibitem[Mathur 2000]{m00} Mathur, S., 2000, MNRAS, 314, L17
\bibitem[McLure \& Jarvis 2002]{mj02} McLure, R.\ J., \&  Jarvis, M.\
  J., 2002, \mnras, 337, 109
\bibitem[Mineshige et al. 2000]{mkth00} Mineshige, S., Kawaguchi, T.,
  Takeuchi, M., Hayashida, K., 2000, PASJ, 52, 499
\bibitem[Murray et al. 1995]{mcgv95} Murray, N., Chiang, J., Grossman,
  S.\ A., \& Voit, G.\ M., 1995, ApJ, 451, 498
\bibitem[Murray \& Chiang 1998]{mc98} Murray, N., \& Chiang, J., 1998,
ApJ, 494, 125
\bibitem[Nicastro, Fiore \& Matt 1999]{nfm99} Nicastro, F., Fiore, F.,
  \& Matt, G., 1999, ApJ, 517, 108
\bibitem[Norris et al. 1990]{naskt90} Norris, R.\ P., Allen, D.\ A.,
Sramek, R.\ A., Kesteven, M.\ J., \& Troup, E.\ R. 1990, \apj, 359, 291
\bibitem[Nussbaumer and Storey (1982)]{nuss82} Nussbaumer, H., \&
Storey, P.\ J., 1982, A\&A, 115, 205
\bibitem[Osterbrock \& Pogge 1985]{op85} Osterbrock, D.\ E., \& Pogge,
  R.\ W., 1985, ApJ, 297, 166
\bibitem[Peterson 1993]{brad93} Peterson, B.\ M., 1993, PASP, 105, 247
\bibitem[Phillips 1976]{p76} Phillips, M.\ M., 1976, ApJ, 208, 37
\bibitem[Pogge \& Owen 1993]{po93} Pogge, R.\ W., \& Owen, J.\ M.,
  1993, Ohio State University Internal report 93-01
\bibitem[Pounds, Done \& Osborne 1995]{pdo95} Pounds, K.\ A., Done,
  C., \& Osborne, J.\ P., 1995, MNRAS, 277, 5P
\bibitem[Pounds, et al. 2001]{pemv01}Pounds, K.\ A., Edelson, R.,
  Markowitz, A., \& Vaughan, S., 2001, ApJL, 550, L15
\bibitem[Proga, Stone \& Kallman 2000]{psk00} Proga, D., Stone, J.\
  M., Kallman, T.\ R., 2000, ApJ, 543, 686
\bibitem[Puchnarewicz et al. 2001]{pms01} Puchnarewicz, E.\ M., Mason,
  K.\ O., Siemiginowska, A., Fruscione, A., Comastri, A., Fiore, F.,
  \& Cagnoni, I., 2001, ApJ, 505, 644
\bibitem[Richards et al.\ 2002]{retal02} Richards, G.\ T., Vanden
  Berk, D.\ E., Reichard, T.\ A., Hall, P.\ B., Schneider, D.\ P.,
  SubbaRao, M., Thakar, A.\ R., \& York, D.\ G., 2002, AJ, 124, 1
\bibitem[Rodri\'iguez-Pascual, Mas-Hesse \& Santos-Lle\'o 1997]{rms97}
  Rodri\'iguez-Pascual, P.\ M. , Mas-Hesse, J.\ M., \&
Santos-Lle\'o, M., 1997, A\&A., 327, 72
\bibitem[Siebert et al. 1999]{sllbbm99}Siebert, J., Leighly, K.\ M.,
Laurent-Muehleisen, S.\ A., Brinkmann, W., Boller, Th., \& Matsuoka,
M., 1999, \aa, 348, 678
\bibitem[Sigut \& Pradhan 2003]{sp03} Sigut, T.\ A.\ A.,  \& Pradhan,
  A.\ K., 2003, ApJS, 145, 15
\bibitem[Smith et al. 1997]{ssah97} Smith, P.\ S., Schmidt, G.\ D.,
  Allen, R.\ G., \& Hines, D.\ C., 1997, ApJ, 488, 202
\bibitem[Turner \& Pounds 1989]{tp89} Turner, T.\ J., \& Pounds, K.\
  A., 1989, MNRAS, 240, 833
\bibitem[Tytler \& Fan 1992]{tf92} Tytler, D., \& Fan, X.-M., 1992,
  ApJS, 79, 1
\bibitem[Vestergaard \& Wilkes 2001]{vw01} Vestergaard, M., \& Wilkes,
  B.\ J., 2001, ApJS, 134, 1
\bibitem[Wandel \& Boller 1998]{wb98} Wandel, A., \& Boller, Th.,
  1998, A\&A, 331, 884
\bibitem[Wilkes et al.\ 1999]{w99} Wilkes, B.\ J., Kuraszkiewicz, J.,
  Green, P.\ J., Mathur, S., \& McDowell, J.\ C., 1999, ApJ, 513, 76
\bibitem[Wills et al.\ 1993]{will94} Wills, B.\ J., Brotherton, M.\
S., Fang, D., Steidel, C.\ C., \& Sargen, W.\ L.\ W., 1993, ApJ, 415, 563
\bibitem[Wills et al.\ 1999]{wills99} Wills, B.\ J., Laor, A.,
  Brotherton, M.\ S., Wills, D., Wilkes, B.\ J., Ferland, G.\ J., \&
  Shang, Z., 1999, ApJL, 515, 53
\bibitem[Zheng et al.\ 1997]{z97} Zheng, W., Kriss, G.\ A.\, Telfer,
  R.\ C.\, Grimes, J.\ P.\, \& Davidson, A.\ F.\ 1997, ApJ, 475, 469

\end{thebibliography}
\end{document}